\documentclass[12pt,a4paper]{article}
\usepackage[a4paper]{geometry}

\usepackage{amsmath}
\usepackage{amsfonts}
\usepackage{amssymb}
\usepackage{dsfont}
\usepackage{arydshln}
\usepackage{graphicx}
\usepackage[nosort]{cite}

\allowdisplaybreaks[2]
\numberwithin{equation}{section}

\newcommand{\eq}[1]{\begin{equation}
                     \begin{split} #1 \end{split}
                     \end{equation}}

\newcommand{\op}{\hspace{1pt}}                     
\newcommand{\ul}{\underline}
\newcommand{\ov}{\overline}

\begin{document}


\vspace*{-1.5cm}

\begin{flushright}
  {\small
  LMU-ASC 09/16 \\
  MPP-2016-24
  }
\end{flushright}

\vspace*{1cm}


\begin{center}
{\LARGE
Towards a world-sheet description of doubled \\[6pt]  geometry in string theory
}
\end{center}


\vspace{0.4cm}

\begin{center}
  Ioannis Bakas$^{\,1}$, Dieter L\"ust$^{\,2,3}$, Erik Plauschinn$^{\,3}$
\end{center}


\vspace{0.4cm}

\begin{center} 
\emph{$^{1\,}$Department of Physics \\
School of  Applied Mathematics and Physical Sciences \\
National Technical University \\ 15780 Athens, Greece}  \\
\vspace{0.5cm}
\emph{$^{2\,}$Max-Planck-Institut f\"ur Physik \\
F\"ohringer Ring 6 \\ 80805 M\"unchen, Germany} \\
\vspace{0.5cm}
\emph{$^{3\,}$Arnold-Sommerfeld-Center f\"ur Theoretische Physik \\
Department f\"ur Physik \\ Ludwig-Maximilians-Universit\"at M\"unchen \\ Theresienstra\ss e 37, 80333 M\"unchen, Germany}
\end{center} 

\vspace{1cm}


\begin{abstract}
\noindent
Starting from a sigma-model for a doubled target-space geometry, 
we show that  the number of target-space dimensions can be reduced by half through a gauging procedure.
We apply this formalism to a class of backgrounds relevant for double field theory, 
and illustrate how choosing different gaugings leads to  string-theory configurations T-dual to each other.
We furthermore discuss that given a conformal doubled theory, the reduced theories are conformal as well.

\noindent
As an example we consider the three-dimensional $SU(2)$ WZW model 
and show that the only possible reduced
backgrounds are the cigar and trumpet CFTs in two dimensions, which 
are indeed T-dual to each other.
\end{abstract}

\clearpage


\tableofcontents


\section{Introduction}

One of string theory's attractive  features is its rich structure of 
dualities \cite{Witten:1995ex}. 
This implies that two seemingly
different formulations of string theory can be physically equivalent. 
An example for such a duality is 
T-duality, where string theory compactified on a
circle of radius $R$ is equivalent to a compactification on a circle of radius $\alpha'/R$. 
Another example is 
S-duality, which relates a weakly-coupled regime 
to a strongly-coupled one. 
Furthermore, the AdS/CFT duality makes a connection between 
a gravity theory on AdS spaces and conformal field theories on their boundaries;
and there are many more duality relations relevant for string theory.

In this paper we are interested in T-duality, for which an extensive review can be found in \cite{Giveon:1994fu}.
More concretely, when compactifying string theory on a circle of radius $R$, the theory can be quantized 
and the spectrum can be determined explicitly. T-duality then means that the spectrum is invariant under
the map $R\rightarrow \alpha'/R$. For curved backgrounds, it is in general not 
known how to perform the quantization, however, the mapping between T-dual configurations is often still possible and is provided by the 
Buscher rules \cite{Buscher:1985kb,Buscher:1987sk,Buscher:1987qj}. Furthermore, 
Ro\v cek and Verlinde  have shown in \cite{Rocek:1991ps} that T-duality is 
not only a symmetry of the spectrum, but
a true symmetry of the underlying conformal field theory. Other related work in this context can be found in 
\cite{Giveon:1988tt,Giveon:1991jj,Giveon:1993ph,Alvarez:1993qi,Alvarez:1994wj,Bakas:1995hc,Plauschinn:2013wta,Plauschinn:2014nha,Chatzistavrakidis:2015lga}.

An interesting application of dualities is that novel solutions with potentially new features can be obtained from 
existing ones. The prime example for this idea in the case of T-duality  is the three-torus with $H$-flux, 
which we briefly review.
Applying a T-duality transformation to this background gives the twisted 
torus \cite{Dasgupta:1999ss,Kachru:2002sk}, which is a geometric space carrying a so-called
geometric flux. 
After a second duality transformation a locally geometric configuration is obtained, which is 
however globally  non-geometric \cite{Hellerman:2002ax}. 
Here, the transition functions between different coordinate patches are no longer only diffeomorphisms, 
but involve duality transformations \cite{Dabholkar:2002sy}.
This construction is called a T-fold \cite{Hull:2004in}, and it carries a so-called $Q$-flux \cite{Shelton:2005cf}. 
It has also been argued that a third T-duality transformation can (at least formally) be performed ending in a space with $R$-flux
\cite{Shelton:2005cf}.
The non-geometric spaces spaces carry non-commutative and non-associative structures and 
have been studied from a  mathematical point of view in 
\cite{Mathai:2004qq,Mathai:2004qc,Bouwknegt:2004ap,Bouwknegt:2004tr}, later in \cite{Ellwood:2006my,Grange:2006es}, and have
been reconsidered in a series of papers \cite{Blumenhagen:2010hj,Lust:2010iy,Blumenhagen:2011ph,Blumenhagen:2011yv,Lust:2012fp,Plauschinn:2012kd,Condeescu:2012sp,Mylonas:2012pg,Chatzistavrakidis:2012qj,Andriot:2012vb,Bakas:2013jwa,
Deser:2013pra,Mylonas:2013jha,Blair:2014kla,Mylonas:2014aga,Aschieri:2015roa}. 
Thus, as mentioned above, applying T-duality transformations to known backgrounds can lead to 
new string-theory configurations with novel features.

There are a number of different approaches to study non-geometric backgrounds in particular, and T-duality in 
general. From the word-sheet point of view, left-right asymmetric string CFT constructions, which have been already constructed several years ago
\cite{Kawai:1986ah,Lerche:1986cx,Antoniadis:1986rn,Narain:1986qm},
correspond to non-geometric string backgrounds.
Further aspects of non-geometric string constructions
have been analyzed from a  world-sheet point of view in 
\cite{Flournoy:2005xe,Halmagyi:2008dr,Halmagyi:2009te,Rennecke:2014sca,Bakas:2015gia,Chatzistavrakidis:2015vka}.
Non-geometric flux configurations  have been investigated via field redefinitions
for the ten-dimensional supergravity action in 
\cite{Andriot:2011uh,Andriot:2012wx,Andriot:2012an,Blumenhagen:2012nk,Blumenhagen:2012nt,Blumenhagen:2013aia,Andriot:2013xca,Andriot:2014uda}.
Another strategy is to make T-duality into a manifest symmetry of an action.
This can be achieved by introducing additional auxiliary, but
unphysical, coordinates.   
Through this procedure the dimension of space-time is doubled -- and the physical configuration is obtained by 
removing half of the coordinates. Different choices for which coordinates are removed then 
lead to backgrounds which are T-dual to each other.
\begin{itemize}

\item From the world-sheet point of view, such a construction has 
first appeared in \cite{Tseytlin:1990nb,Tseytlin:1990va}, based on work in \cite{Duff:1989tf}, and
has been revisited for instance in \cite{Dall'Agata:2008qz} (see also \cite{Hohm:2013jaa}).
In this formulation manifest Lorentz invariance is however broken, which has later 
been addressed in \cite{Nibbelink:2012jb,Nibbelink:2013zda}.
In \cite{Hull:2004in,Dabholkar:2005ve,Hull:2006va} 
a different doubled formalism appeared, 
in which  a constraint to reduce the number of coordinates has to be imposed by hand.

\item From a target-space perspective, an action invariant under T-duality has been developed 
(using also world-sheet techniques)
in \cite{Siegel:1993xq,Siegel:1993th}, but also here at the cost of broken Lorentz covariance. 
In \cite{Hull:2009mi,Hohm:2010jy,Hohm:2010pp},
combining insight from string field theory with the doubled formalism,
the framework of double field theory (DFT) has been developed 
(for reviews see \cite{Aldazabal:2013sca,Hohm:2013bwa}).
Double field theory is a field theory with a doubled number of target-space dimensions, which is manifestly 
invariant under T-duality. Though, on the other hand, it is in general not known how to incorporate massive excitations 
of the string into this setting.

\item We also mention that from a mathematical point of view, the idea to double the coordinates 
in order to make T-duality a symmetry has been 
discussed in \cite{Hori:1999me,Bouwknegt:2003vb}.

\end{itemize}

In this paper, we follow the approach of doubling the number of target-space dimensions in order to 
make T-duality a manifest symmetry. We consider a world-sheet formulation, and we discuss the conformal properties of the corresponding doubled and physical theories. 
Our main strategy is the following:
\begin{enumerate}

\item We start from a background in $D$ dimensions with two directions of isometry. 
The background is required to have vanishing $\beta$-functionals for the metric and $B$-field,
but the $\beta$-functional for the dilaton is non-zero at tree-level.
It is therefore not a critical string theory.

\item Next, we gauge one of the two isometries in the world-sheet theory, and integrate-out 
the corresponding gauge field. As we show in detail, the resulting metric and $B$-field 
are that of a $(D-1)$-dimensional target-space theory. 
The reduced theory in $D-1$ dimensions is conformal, in particular it is a string theory.

\item Gauging a different isometry leads to a different $D$-dimensional background.
However, these lower-dimensional theories are related to each other through 
T-duality transformations.

\end{enumerate}
Thus, we provide a world-sheet approach to doubled geometry, in which the 
T-dual backgrounds originate from different choices in a reduction procedure.
The novel aspect of our analysis is that the reduction is achieved by gauging isometries
and integrating-out the corresponding gauge field, instead of imposing 
constraints explicitly.

We also would like to point out relations between our work and other approaches. 
First, our strategy is closely related to Buscher's procedure
of gauging an isometry and integrating-out the gauge field $A$ \cite{Buscher:1985kb,Buscher:1987sk,Buscher:1987qj}. 
However, in Buscher's work a Lagrange multiplier is introduced which 
implements the vanishing of the  field strength $F=dA$. 
This Lagrange multiplier becomes the 
dual coordinate after T-duality, and hence the dimensionality of the target space is unchanged. 
In our work, we impose the vanishing of the field strength by hand, without including a Lagrange
multiplier. We therefore find that the number of space-time dimensions is reduced. 
Second, there is no direct connection to the canonical-transformation approach
to T-duality discussed for instance in \cite{Giveon:1988tt,Alvarez:1994wj}, since we obtain a reduction of the 
dimensionality of the target-space.
Third, similar to Buscher's approach, a Killing vector field $k$ and a one-form $\mathsf v$
will play a prominent role in our analysis. 
In particular, T-duality in the reduced theories will be realized by
interchanging the role of $k$ and $\mathsf v$. 
Such a transformation can be described using the framework of generalized geometry 
(see \cite{Hitchin:2004ut} and \cite{Gualtieri:2003dx} for the original work,
and for instance to \cite{Grana:2008yw} for a discussion in the context of T-duality), in which the tangent and co-tangent space are treated on equal footing. 
Finally, we mention that a related world-sheet discussion has appeared in 
\cite{Nibbelink:2012jb,Nibbelink:2013zda}.

\bigskip
This paper is organized as follows: in section~\ref{sec_dim_red} we discuss 
how by gauging certain symmetries of a non-linear sigma-model and integrating-out
the corresponding gauge field, the number of target-space dimensions is reduced. 
We first consider the case of a single doubled dimension, and then 
generalize to multiple doubled dimensions. 
In section~\ref{sec_rv}, we apply this procedure to a particular type of backgrounds
relevant for double field theory. These backgrounds have also appeared in the work of
Ro\v cek and Verlinde \cite{Rocek:1991ps}, 
and we find that our results incorporate those of \cite{Rocek:1991ps} nicely. 
In section~\ref{sec_ex} we first discuss the $SU(2)$ Wess-Zumino-Witten model as an example, 
and then comment on the generalization to arbitrary Lie groups. 
Section~\ref{sec_sc} contains our summary and conclusions.


\section{Dimensional reduction by gauging isometries}
\label{sec_dim_red}

In this section we discuss on general grounds how 
by gauging a symmetry of a non-linear sigma-model and integrating-out the gauge field, 
the number of target-space dimensions
can be reduced.


\subsection{Gauging an isometry}
\label{sec_gauge_iso}

Let us start by reviewing the gauging procedure for non-linear sigma-models. 
This has been discussed by Hull and Spence in \cite{Hull:1989jk,Hull:1990ms},
and has recently been revisited and further analyzed in  
\cite{Plauschinn:2013wta,Plauschinn:2014nha}.


\subsubsection*{Sigma-model action and isometries}

Our starting point is the 
a non-linear sigma-model action for a metric $G$, anti-symmetric tensor field $B$ and a dilaton $\Phi$.
We employ a formulation of this action as a Wess-Zumino-Witten (WZW) model, where instead of $B$ itself the field strength $H=dB$
appears. The action reads
\eq{
  \label{action_01}
  \mathcal S =& -\frac{1}{4\pi \alpha'} \int_{\partial\Sigma} 
  \Bigl[ G_{ij} \, d X^i\wedge\star d X^j
  + \alpha'R\, \Phi \star 1 \Bigr] \\[2mm]
  &-\frac{i}{2\pi \alpha'} \int_{\Sigma} \tfrac{1}{3!}\, H_{ijk}\op dX^i\wedge dX^j\wedge dX^k
  \,,
}
where the indices $i,j,k=1,\ldots, D$ label the target-space coordinates. 
The Hodge star operator on the world-sheet is denoted by $\star$,  $\Sigma$ is a
three-manifold with boundary $\partial\Sigma$, and $R$ denotes the 
Ricci curvature scalar on the world-sheet $\partial\Sigma$.

In order to follow the procedure we have in mind, we assume the 
action \eqref{action_01} to be invariant
under the following global variation
\eq{
  \label{iso_trafo_01}
  \hspace{50pt}
  \delta_{\epsilon} X^i = \epsilon\op k^i(X) \,,
  \hspace{70pt} \epsilon={\rm const.}
}
This requirement translates into three conditions for the target-space metric $G$, the field strength $H$ and the dilaton $\Phi$,
which read as follows:
\begin{itemize}

\item First, the vector $k = k^i\op \partial_i$ has to be a Killing vector for the metric $G$.
Using the coordinate-free notation $G=G_{ij}\op dX^i\wedge \star dX^j$ and the Lie derivative
$\mathcal L_k=d\circ \iota_k + \iota_k\circ d$, this reads
\eq{
 \label{iso_01}
  \mathcal L_k \op G = 0 \,.
}

\item Second, invariance under global variations \eqref{iso_trafo_01} implies for the field strength 
\eq{
 \label{iso_02}
  \iota_k H = d \op \mathsf v \,,
}
for $\mathsf v$ a one-form on the target space. Note that  
$\mathsf v$ in \eqref{iso_02} is defined only up to a closed part.
Equation \eqref{iso_02} is equivalent to 
$\mathcal L_k  B= d\op \mathsf v'$ with $\mathsf v'$ some other one-form, 
which then leads to
\eq{
 \label{iso_04}
 \mathcal L_k  H= 0 \,.
}

\item The third condition for the variation of the action to vanish is 
\eq{ 
  \label{iso_05}
  \mathcal L_k \Phi = k^m \partial_m \Phi = 0 \,.
}

\end{itemize}


\subsubsection*{Gauging the sigma-model}

Next, we gauge the symmetry \eqref{iso_trafo_01}
by allowing $\epsilon$ to have a non-trivial dependence 
on the world-sheet coordinates.  We therefore  introduce a gauge field $A$ and 
replace $dX^i\to dX^i + k^i A$ for the term
involving the metric. For the Wess-Zumino term we keep $dX^i$ unchanged, but introduce a
coupling between the one-form $\mathsf v$ and the gauge field $A$.
The resulting gauge-invariant action  takes the following form \cite{Hull:1989jk,Hull:1990ms}
\begin{align}
  \nonumber
  \widehat{\mathcal S} =&-\frac{1}{4\pi\alpha'} \int_{\partial\Sigma} \Bigl[  
  G_{ij}  (dX^i + k^i A)\wedge\star(dX^j + k^j A)  
  +2\op i \op\mathsf  v\wedge A
  + \alpha'R\, \Phi \star 1 \Bigr] \\[2mm]
  \label{action_02}
  &-\frac{i}{2\pi \alpha'} \int_{\Sigma}  \tfrac{1}{3!}\, H_{ijk}\op dX^i\wedge dX^j\wedge dX^k\,,
\end{align}
where $k$ denotes again the Killing vector of the target-space isometry which has been gauged. 
The symmetry transformations for the fields in the action read
\eq{
  \label{iso_trafo_02}
  \hat\delta_{\epsilon} X^i = \epsilon\op k^i(X) \,, \hspace{70pt} 
  \hat\delta_{\epsilon} A = - d \op\epsilon\,.
}
Moreover, for gauge invariance of the gauged action \eqref{action_02} we  have to require \cite{Hull:1989jk,Hull:1990ms}
\eq{
  \label{req_03}
  \iota_k \mathsf v =  k^m \mathsf v_m  = 0 \,.
}

Note that by gauging the symmetry \eqref{iso_trafo_01} we have introduced
two new degrees of freedom into the theory, out of which one can be eliminated by a gauge transformation
\eqref{iso_trafo_02}.
In order to eliminate the second additional degree of freedom, and 
for the gauged action \eqref{action_02} to be 
equivalent to the ungauged one \eqref{action_01}, we impose
the constraint 
\eq{
  \label{con_007}
  0=F = dA \,.
}
When studying T-duality, this constraint is usually realized using a Lagrange multiplier. However, here
we chose to impose the constraint by hand.


\subsection{Reduction of dimensions}

After gauging an isometry of world-sheet action, we now want to integrate-out the 
corresponding gauge field. This leads to a reduction of the number of target-space dimensions.


\subsubsection*{Integrating-out the gauge field}

To integrate-out the gauge field $A$,  we determine its equation of motion following from  \eqref{action_02}. 
With $|k|^2=k^i G_{ij} k^j$ the norm of the Killing vector field, we 
find
\eq{
  \label{eom_A}
 |k|^2 A = -  k^i G_{ij}\op dX^j - i \star \mathsf v  \,.
}
If  $|k|^2$  is non-vanishing, we can solve \eqref{eom_A}
for $A$ and substitute the solution back into the action. 
The resulting world-sheet action takes the general form
\eq{
  \label{action_05}
  \check{\mathcal S} =  
  &-\frac{1}{4\pi \alpha'} \int_{\partial\Sigma} \Bigl[ \check G_{ij} \, 
   d X^{i}\wedge\star d X^{j}
  + \alpha'R\, \Phi \star 1 \Bigr] \\[2mm]
  &-\frac{i}{2\pi \alpha'} \int_{\Sigma} \tfrac{1}{3!}\, \check H_{ijk}\, 
  dX^{i}\wedge dX^{j}\wedge dX^{k} \,.
}
With $\mathsf k = k^i\op G_{ij} \op dx^j$ the one-form dual to the Killing vector $k$, 
we find for $\check G$ and  $\check H$ the expressions
\eq{
  \label{g_and_h_01}
  \check G = G - \frac{1}{|k|^2}\, \mathsf k \wedge \star \mathsf k +
   \frac{1}{|k|^2}\, \mathsf v \wedge\star \mathsf v\,,
   \hspace{40pt}
  \check H = H 
  + d\left( \frac{1}{|k|^2}\, \mathsf k \wedge \mathsf v\right)
   \,.
}
Let us also recall that the gauge field $A$ is subject to the constraint \eqref{con_007},
in particular, the solution \eqref{eom_A} has to satisfy $dA=0$. 
Using the equations of motion for $X^i$, 
we can express this constraint as\op\footnote{Our convention is that the symmetrization and
anti-symmetrization of indices contains a factor of $1/n!$.}
\eq{
  \label{more_cons}
   0 = \nabla_{[ \op \ul i} \op \mathsf k_{\ul j \op]} - \frac{1}{2} \op v^m H_{mij} \,, 
   \hspace{50pt}
   0 = \nabla_{( \op \ov i} \op \mathsf v_{\ov j \op)}   \,,
   \hspace{50pt}
   v^m \partial_m \Phi = 0 \,,
}  
where the Levi-Civita connection appearing in $\nabla$ is computed using the original metric $G$,
and $v^i = G^{ij} \mathsf v_j$ are the components of the vector field $v=v^i\partial_i$ 
dual to the one-form $\mathsf v$.
Details about the derivation of these relations can be found in appendix~\ref{app_da}.


\subsubsection*{Reduced geometry}

Next, we observe  that $\check G$  in \eqref{g_and_h_01}
has an eigenvector with zero eigenvalue. Indeed, let us 
consider
\eq{
  \label{lie_enl_02}
  \iota_k \check G = 2 \op k^i G_{ij} \star dX^j - 2\star \mathsf k + \frac{2}{|k|^2} \, (\iota_k \mathsf v) \star \mathsf v
  = 0 \,,
}
which vanishes due to \eqref{req_03}. In a similar way, for the field strength $\check H$ we find after
a short computation that
\eq{
  \label{lie_enl_03}
  \iota_k \check H = 0 \,.
}
Thus, even though the original $D$-dimensional metric $G_{ij}$ is 
non-degenerate, 
the matrix $\check G_{ij}$ defined via  \eqref{g_and_h_01} 
has one vanishing eigenvalue with eigenvector 
$k$. Moreover, it turns out that $k$ is also a Killing vector for $\check G$ (and similarly for $\check H$), that is
\eq{
  \label{lie_enl_01}
  \mathcal L_k \check G = 0 \,, \hspace{80pt}
  \mathcal L_k \check H = 0\,.
}

Given that $k$ is a null-vector, we can perform a change of 
coordinates. Assuming without loss of generality that $k^1$ is non-zero, we transform the matrix $\check G$ as
\eq{
  \label{coc_01}
  \check{\mathcal G}_{ij} =  \bigl( \mathcal T^T \check G \,\mathcal T 
  \bigr)_{ij}\,,
  \hspace{70pt}
  \mathcal T^i{}_j=
  \scalebox{0.9}{$\displaystyle
  \renewcommand{\arraystretch}{1.35}
  \arraycolsep6pt
  \dashlinedash2pt
  \dashlinegap4pt  
  \left( \begin{array}{c:c@{\hspace{17pt}}c@{\hspace{17pt}}c}
  k^1 & \multicolumn{3}{c}{0} \\ \hdashline
  k^2 &  &&    \\[-2pt]
  \vdots & & \mathds 1 & \\[-2pt]
  k^D & &&    
  \end{array}
  \right)
  $}.
}
In the transformed matrix $\check{\mathcal G}_{ij}$ 
all entries along the $i,j=1$ direction vanish, and 
we therefore arrive at
the  expression
\eq{
  \label{coc_02}
  \check{\mathcal G}_{ij} =
  \renewcommand{\arraystretch}{1.4}
  \arraycolsep6pt
  \dashlinedash2pt
  \dashlinegap4pt  
  \left( \begin{array}{c:c@{\hspace{2pt}}c@{\hspace{2pt}}c}
  0 & \multicolumn{3}{c}{0}  \\ \hdashline
  &  &&     \\[-12pt]
  0 & & \,\check G_{ab} & \\[-12pt]
  & &&    \\ 
  \end{array}
  \right) ,
}
where $a,b=2,\ldots,D$.
Turning to the field strength $\check H$ and employing
the matrix $\mathcal T^i{}_j$, we transform $\check H$ as follows
\eq{
  \check{\mathcal H}_{ijk} = \check H_{lmn} \mathcal T^l{}_i\mathcal T^m{}_j\mathcal T^n{}_k\,.
}  
Similarly to the transformed metric $\check{\mathcal G}_{ij}$, we again find that all 
components of $\check{\mathcal H}$ 
along the $i=1$ direction vanish, that is
\eq{
  \label{coc_13}
  \check{\mathcal H}_{1 jk } = 0 \,.
}
From  \eqref{coc_02} and \eqref{coc_13} we can conclude that in the action
\eqref{action_05} the forms corresponding to the $i=1$ direction have dropped out.


\subsubsection*{Change of coordinates}

However, this observation does not imply that we have arrived at a lower-di\-men\-sional theory.
In particular, let us  consider the transformed basis of one-forms given 
by $e^i = (\mathcal T^{-1} )^i{}_j \op dX^j$.
For the transformation matrix $\mathcal T$ shown in \eqref{coc_01} we find 
\eq{
  \label{coc_11}
  e^1 = \frac{1}{k^1}\, dX^1\,, \hspace{70pt}
  e^{a} = dx^{a}- \frac{k^{a}}{k^1}\, dX^1 \,, 
}
where again $a=2,\ldots, D$.
Note that the algebra of one-forms $\{ e^{a}\}$ in general does not close, but 
requires the basis of forms in the full $D$-dimensional space. More concretely, we have
\eq{
  de^{a} & = 
   \frac{1}{k^1}\op\Bigl(  k^1 \partial_{b} k^{a} -k^{a}\partial_{b} k^1 \Bigr)\,
  e^1\wedge e^{b} \,.
}
Therefore, in order for $\{ e^{a}\}$ to close on itself and to properly reduce the $D$-dimensional 
target space to  $(D-1)$ dimensions,
for all $a,b\in\{2,\ldots, D\}$ we  have to require 
\eq{ 
  \label{coc_04}
  0= k^1 \partial_{b} k^{a} -k^{a}\partial_{b} k^1 \,.
 }

Performing a change of coordinates in the original geometry, 
it is always possible to choose 
$k^{a}=0$ for all $a=2,\ldots, D$. The requirement \eqref{coc_04} 
can therefore be satisfied. 
Furthermore, using \eqref{lie_enl_01} and \eqref{iso_05} together with $k^a=0$
we can show that the components $\check G_{ab}$ and $\check H_{abc}$
do not depend on the coordinates in the direction $k$. 
Hence, after this change of coordinates we have 
arrived at a $(D-1)$-dimensional target-space background.


\subsection{Generalization to multiple isometries}

In the last two sections, we have described how to reduce the number of target-space dimensions 
of a non-linear sigma-model by gauging a single world-sheet symmetry. 
In this section, following \cite{Plauschinn:2014nha}, we generalize this procedure
to multiple, possibly non-abelian, symmetries.


\subsubsection*{Original action and isometries}

We start again from the ungauged sigma-model action shown in equation \eqref{action_01},
and assume that this action is invariant under the global symmetries
\eq{
  \label{iso_trafo_01a}
    \delta_{\epsilon} X^i = \epsilon^{\alpha}\op k_{\alpha}^i(X) \,,
}
labeled by $\alpha = 1, \ldots, N$. 
We can assume without loss of generality that the vectors $k_{\alpha}$ 
are linearly independent.
Furthermore, in order for the algebra of variations to close, the algebra generated by the vectors $k_{\alpha}$ 
is required to close, 
\eq{
   \bigl[ k_{\alpha} , k_{\beta} \bigr]_{\rm L} = f_{\alpha\beta}{}^{\gamma} \op k_{\gamma} \,.
}
Demanding then the action \eqref{action_01} to be invariant
under \eqref{iso_trafo_01a} translates into the following conditions, 
generalizing the expressions given in section~\ref{sec_gauge_iso}
\eq{
  \label{req_001}
  \mathcal L_{k_{\alpha}}  G = 0\,, \hspace{70pt}
  \iota_{k_{\alpha}} H = d\mathsf v_{\alpha} \,, \hspace{70pt}
  \mathcal L_{k_{\alpha}} \Phi =0\,.
}
The one-forms $\mathsf v_{\alpha}$ are defined only up to a closed part, 
and their dual vector fields will be denoted by $v_{\alpha}$ in the following.


\subsubsection*{Gauged action}

Next, we gauge the symmetries \eqref{iso_trafo_01a}. The resulting action has
 been derived in \cite{Hull:1989jk,Hull:1990ms} and reads (in the notation of 
\cite{Plauschinn:2014nha}) as 
\eq{
  \label{action_021}
  \widehat{\mathcal S} =&-\frac{1}{2\pi\alpha'} \int_{\partial\Sigma} \:
  \Bigl[ \;
  \tfrac{1}{2}\op G_{ij}  (dX^i + k^i_{\alpha} A^{\alpha})\wedge\star(dX^j + k^j_{\beta} A^{\beta})  
  + \tfrac{\alpha'}{2}  R\, \Phi \star 1 \;\Bigr]
  \\[1mm]
  &-\frac{i}{2\pi \alpha'} \int_{\Sigma} \hspace{13.5pt} \tfrac{1}{3!}\, H_{ijk}\op dX^i\wedge dX^j\wedge dX^k
  \\[1mm]
  &-\frac{i}{2\pi \alpha'} \int_{\partial\Sigma} \:\Bigl[ \;
   \mathsf v_{\alpha} \wedge A^{\alpha}
  + \tfrac{1}{2}\op \bigl( \iota_{k_{[\ul \alpha}} \mathsf v_{\ul \beta]} \bigr)
  \op A^{\alpha}\wedge A^{\beta}\;
  \Bigr] \,.
}
The local symmetry transformations take the form
\eq{
  \label{variantions_02a}
  \hat\delta_{\epsilon} X^i = \epsilon^{\alpha} \op k^i_{\alpha} \,,\hspace{50pt}
  \hat \delta_{\epsilon} A^{\alpha} = - d\epsilon^{\alpha} - \epsilon^{\beta} A^{\gamma} f_{\beta\gamma}{}^{\alpha}
  \,,
}
and for invariance of the gauged action \eqref{action_021} under the transformations
\eqref{variantions_02a} we have  to require that
\eq{
  \label{req_002}
  \mathcal L_{k_{[\ul \alpha}} \mathsf v_{\ul \beta]} = f_{\alpha\beta}{}^{\gamma} \mathsf v_{\gamma} \,,
  \hspace{70pt}
  \iota_{k_{(\ov \alpha}} \mathsf v_{\ov \beta)} =0\,.
}
Finally, we mention that by introducing the gauge fields $A^{\alpha}$, we again have 
enlarged the number of degrees of freedom of the theory. In order for the gauged action 
to be equivalent to the ungauged one, we demand the field strengths $F^{\alpha}$ to vanish, 
that is
\eq{
  \label{da_is_zero}
  0 = F^{\alpha} = d A^{\alpha} -\tfrac{1}{2}\op f_{\beta\gamma}{}^{\alpha} A^{\beta}\wedge A^{\gamma}  \,.
}


\subsubsection*{Integrating-out the gauge field}

The next step is to integrate-out the gauge field. The equations of motion 
for $A^{\alpha}$ read
\eq{
  \label{eom_30}
  A^{\alpha} = - \bigl( \mathcal M^{-1} \bigr)^{\alpha\beta} \Bigl(
  \mathds 1 + i \star \mathcal D \, \mathcal G^{-1} \Bigr)_{\beta}^{\;\;\gamma} \bigl(\, 
  \mathsf k + i \star \mathsf v \bigr)_{\gamma}\,,
}
where we remind the reader that $\alpha,\beta,\gamma=1,\ldots,N$ label the isometries which have been gauged.
In the above expression, we have employed the notation
\eq{
  \label{back_67}
  \arraycolsep2pt
  \begin{array}{lclclcl}
  \mathcal G_{\alpha\beta} &=& k_{\alpha}^i G_{ij} k^j_{\beta} \,, 
  \\[4mm]
  \mathcal D_{\alpha\beta} &=&  \iota_{k_{[\ul \alpha}} v_{\ul \beta]} \,,
  &\hspace{50pt}&
  \mathsf k_{\alpha} & = & k^i_{\alpha} G_{ij} dX^j \,,
  \\[4.5mm]
  \mathcal M_{\alpha\beta} &=& \bigl(  \mathcal G  - \mathcal D \, \mathcal G^{-1} \mathcal D \bigr)_{\alpha\beta} \,,
  \end{array}
}
and have assumed the matrices $\mathcal G$ and $\mathcal M$ to be invertible,
\eq{
  \label{invert_067}
  \det \mathcal G \neq 0 \,, \hspace{70pt}
  \det \mathcal M \neq 0 \,.
}   
In the case of a single Killing vector this corresponds to the usual requirement that
$|k|^2 \neq0$.
After integrating-out the gauge field, the resulting action takes the general form shown in 
\eqref{action_05}. Using matrix multiplication and suppressing the indices $\alpha,\beta,\ldots$
the metric $\check G$ and field strength $\check H$ are given by
\eq{
  \label{red_geo}
  \check G &= G + 
  \binom{\mathsf k}{\mathsf v}^T \hspace{-2.3pt}\left(\begin{matrix} -\mathcal M^{-1} 
  & -\mathcal M^{-1} \mathcal D \op\mathcal G^{-1} \\ 
  +\mathcal M^{-1} \mathcal D \op\mathcal G^{-1}  & +\mathcal M^{-1} \end{matrix} \right)
  \wedge \star\binom{\mathsf k}{\mathsf v} \,,
  \\[3mm]
  \check H &= H + 
  \tfrac{1}{2}  \op d\left[  \binom{\mathsf k}{\mathsf v}^T \left(\begin{matrix} 
   +\mathcal M^{-1} \mathcal D \op\mathcal G^{-1} 
  &+\mathcal M^{-1} \\ 
   -\mathcal M^{-1} & -\mathcal M^{-1} \mathcal D \op\mathcal G^{-1} \end{matrix} \right)
  \wedge \binom{\mathsf k}{\mathsf v} \right] \,.
}
Furthermore, we note that the requirement of vanishing field strength shown in equation \eqref{da_is_zero} 
imposes additional constraints. These are the generalizations of \eqref{more_cons}
which read
\eq{
  \label{more_cons_a}
   0 = \nabla_{[ \op \ul i} \op \mathsf k_{\alpha\op \ul j \op]} - \frac{1}{2} \op v_{\alpha}^m H_{mij} \,, 
   \hspace{50pt}
   0 = \nabla_{( \op \ov i} \op \mathsf v_{\alpha\op \ov j \op)}   \,,
   \hspace{50pt}
   v_{\alpha}^m \partial_m \Phi = 0 \,,
}  
where $v_{\alpha} = \mathsf v_{\alpha\op i} \op G^{ij} \partial_j$ are the vector fields dual to the one-forms
$\mathsf v_{\alpha}$.
Details on the derivation of \eqref{more_cons_a} can be found in appendix~\ref{app_da}.
Moreover, employing  \eqref{req_001} and \eqref{req_002}
we can determine the following algebra
\eq{
   \bigl[ k_{\alpha} , k_{\beta} \bigr]_{\rm L} = f_{\alpha\beta}{}^{\gamma} \op k_{\gamma} \,,
   \hspace{30pt}
   \bigl[ k_{\alpha} , v_{\beta} \bigr]_{\rm L} = f_{\alpha\beta}{}^{\gamma} \op v_{\gamma} \,,
   \hspace{30pt}
   \bigl[ v_{\alpha} ,  v_{\beta} \bigr]_{\rm L} = f_{\alpha\beta}{}^{\gamma} \op k_{\gamma} \,.   
}


\subsubsection*{Reduced geometry}

Next, we observe that, as before, the vectors $k_{\alpha}$ are null-vectors as well as Killing vectors 
for $\check G$ and $\check H$. In particular, 
we have 
\eq{
  \label{iso_enl}
  \arraycolsep2pt
  \begin{array}{lcl@{\hspace{80pt}}lcl}
  \iota_{k_{\alpha}} \check G &=& 0 \,,
  & \mathcal L_{k_{\alpha}} \check G &=& 0 \,, \\[6pt]
  \iota_{k_{\alpha}} \check H &=& 0\,, 
  &\mathcal L_{k_{\alpha}} \check H &=& 0 \,.
  \end{array}
}
We can therefore perform a change of basis in the following way
\eq{
  \label{coc_01a}
  \check{\mathcal G}_{ij} =  \bigl( \mathcal T^T \check G \,\mathcal T 
  \bigr)_{ij}\,,
  \hspace{70pt}
  \mathcal T^i{}_j=
  \scalebox{0.8}{$\displaystyle
  \renewcommand{\arraystretch}{1.35}
  \arraycolsep6pt
  \dashlinedash2pt
  \dashlinegap4pt  
  \left( \begin{array}{ccc:c@{\hspace{17pt}}c@{\hspace{17pt}}c}
  k_1^1  & \cdots & k_N^1 &  \\
  \vdots && \vdots & \multicolumn{3}{c}{0} \\
  k_1^N & \cdots & k_N^N  &  &&    \\[3pt] \hdashline
  \vdots && \vdots & & \raisebox{-10pt}{$\mathds 1$} & \\[-7pt]
  k_1^D & \cdots & k_N^D & &&    
  \end{array}
  \right)
  $},
}
where we arranged the matrix $\mathcal T$ in such a way that the upper-left block is 
invertible. 
(This is always possible provided the Killing vectors are linearly independent.)
The schematic form of the transformed matrix $\check{\mathcal G}$ is
\eq{
  \label{coc_02a}
  \check{\mathcal G}_{ij} =
  \renewcommand{\arraystretch}{1.4}
  \arraycolsep6pt
  \dashlinedash2pt
  \dashlinegap4pt  
  \left( \begin{array}{c:c@{\hspace{2pt}}c@{\hspace{2pt}}c}
  0 & \multicolumn{3}{c}{0}  \\ \hdashline
  &  &&     \\[-12pt]
  0 & & \,\check G_{ab} & \\[-12pt]
  & &&    \\ 
  \end{array}
  \right) ,
}
where $a,b=(N+1),\ldots,D$. Similarly, for the field strength we consider
the transformation $\check{\mathcal H}_{ijk} = \check H_{lmn} \mathcal T^l{}_i\mathcal T^m{}_j\mathcal T^n{}_k$,
for which we find
\eq{
  \check{\mathcal H}_{1 ij} = 0 \,, \hspace{30pt} 
  \check{\mathcal H}_{2 ij} = 0 \,,
  \hspace{30pt}\ldots\hspace{30pt}  
  \check{\mathcal H}_{N ij} = 0\,.
}
Performing furthermore a change of coordinates 
such that $k_{\alpha}^i=0$ for $i=N+1,\ldots, D$ and using \eqref{iso_enl},
we can show that the 
algebra of transformed one-forms $\{e^a\}$ closes under $d$, and that
components $\check G_{ab}$ and $\check H_{abc}$ do not depend
on $X^i$ for $i=1,\ldots, N$. 
Hence, we have arrived at sigma-model whose target-space geometry is reduced by $N$ dimensions.


\subsection{Summary and discussion}


\subsubsection*{Summary}

Let us  briefly summarize and discuss the results obtained in this section. 
\begin{itemize}

\item Our starting point was the non-linear sigma-model action \eqref{action_01}
which we assumed to be invariant  under global transformations 
$\delta_{\epsilon} X^i = \epsilon^{\alpha}\op k_{\alpha}^i$
with $\alpha=1,\ldots, N$. This requirement  implies that the vectors $k_{\alpha}$ have to be Killing, 
it gave rise to the definition of one-forms $\mathsf v_{\alpha}$, and
it implied that $k_{\alpha}^m\partial_m \Phi=0$. In formulas this reads
\eq{
  \label{sum_001}
  \mathcal L_{k_{\alpha}}  G = 0\,, \hspace{60pt}
  \iota_{k_{\alpha}} H = d\op\mathsf v_{\alpha} \,, \hspace{60pt}
  \mathcal L_{k_{\alpha}} \Phi =0\,.
}
Note furthermore that the algebra of Killing vectors can be non-abelian
with structure constants $f_{\alpha\beta}{}^{\gamma}$.

\item Next, we gauged this world-sheet symmetry. The gauged action has been shown
in \eqref{action_021}, and for gauge invariance one has to require \eqref{req_002}, 
that is 
\eq{
  \mathcal L_{k_{[\ul \alpha}} \mathsf v_{\ul \beta]} = f_{\alpha\beta}{}^{\gamma} \mathsf v_{\gamma} \,,
  \hspace{70pt}
  \iota_{k_{(\ov \alpha}} \mathsf v_{\ov \beta)} =0\,.
}

\item Finally, through the gauging procedure gauge fields $A^{\alpha}$  have been introduced. In order 
to for the gauged and ungauged theory to be equivalent, we required that the corresponding 
field strengths vanish, $F^{\alpha}=0$. The latter condition is satisfied provided that
\eq{
  \label{sum_002}
  \mathcal L_{v_{\alpha}}  G = 0\,, \hspace{70pt}
  \iota_{v_{\alpha}} H = d\op\mathsf k_{\alpha} \,, \hspace{70pt}
  \mathcal L_{v_{\alpha}} \Phi =0\,.
}

\end{itemize}
If these conditions are met, it is possible to gauge the isometries associated to 
$k_{\alpha}$ without introducing additional degrees of freedom. 
Integrating-out the gauge field from the action and performing a change 
of coordinates gives a target-space background,
which is reduced by $N$ dimensions. 
The expressions for the reduced metric and field strength are obtained from \eqref{red_geo}
using the change of coordinates shown in \eqref{coc_01a}.


\subsubsection*{Remarks}

Let us conclude this section with the following remarks:
\begin{itemize}

\item We observe that the conditions \eqref{sum_001} and \eqref{sum_002} are the 
same expressions with $k_{\alpha}$ and $v_{\alpha}$ interchanged. Writing them 
in a coordinate-dependent way and after a slight rearrangement, we have 
\begin{align}
  \label{cond_sum}
  \begin{array}[c]{@{}l@{\hspace{40pt}}l@{\hspace{40pt}}l@{}}
   \displaystyle  0 = \nabla_{( \op \ov i} \op \mathsf k_{\alpha\op\ov j \op)}   \,, 
   &
   \displaystyle 0 = \nabla_{[ \op \ul i} \op \mathsf k_{\alpha \op \ul j \op]} - \frac{1}{2} \op v_{\alpha}^m H_{mij} \,, 
   &
   \displaystyle 0 = k_{\alpha}^m \partial_m\Phi \,,
   \\[14pt]
   \displaystyle  0 = \nabla_{( \op \ov i} \op \mathsf v_{\alpha\op \ov j \op)}   \,,
   &
   \displaystyle 0 = \nabla_{[ \op \ul i} \op \mathsf v_{\alpha\op \ul j \op]} - \frac{1}{2} \op k_{\alpha}^m H_{mij} \,,
   &
   \displaystyle 0 = v_{\alpha}^m \partial_m\Phi \,,
   \end{array}
\end{align}  
together with
\eq{
  \label{cond_sum_2}
  0= \iota_{k_{\alpha}} \mathsf v_{ \beta}+ \iota_{k_{\beta}} \mathsf v_{ \alpha} \,,
  \hspace{70pt}
  \arraycolsep2pt
  \begin{array}[c]{ccl}
  \displaystyle \bigl[ k_{\alpha} , k_{\beta} \bigr]_{\rm L}  &=& f_{\alpha\beta}{}^{\gamma} \op k_{\gamma} \,,
  \\[8pt]
  \displaystyle \bigl[ k_{\alpha} , v_{\beta} \bigr]_{\rm L}  &=& f_{\alpha\beta}{}^{\gamma} \op v_{\gamma} \,,
  \\[8pt]
  \displaystyle \bigl[ v_{\alpha} ,  v_{\beta} \bigr]_{\rm L} &=& f_{\alpha\beta}{}^{\gamma} \op k_{\gamma} \,.   
  \end{array}
}

\item We also remark that for vanishing $H$-flux, the components $k_{\alpha}^i$ and $v_{\alpha}^i$ have to be
covariantly constant.
Hence the structure constants of the isometry algebra vanish,
\eq{
  H=0 \hspace{30pt}\Longrightarrow\hspace{30pt}
  \begin{array}{l}
  0 = \nabla_i \,k_{\alpha}^j  \\[6pt]
  0 = \nabla_i \,v_{\alpha}^j 
  \end{array}
  \hspace{30pt}\Longrightarrow\hspace{30pt}
  f_{\alpha\beta}{}^{\gamma}=0\,.
}
Furthermore, for $H=0$ the one-forms $\mathsf v_{\alpha}$, 
 and consequently the vector fields $v_{\alpha}$,
can all be chosen to be zero.

\end{itemize}


\section{Relation to Ro\v cek \& Verlinde}
\label{sec_rv}

In this section we make contact between our discussion in section~\ref{sec_dim_red}
and the work of Ro\v cek and Verlinde (RV) \cite{Rocek:1991ps}, which 
has also been used in \cite{Bakas:2015gia}. In particular, we assume the metric 
$G$ and the field strength $H$ appearing in the world-sheet 
action \eqref{action_01} to have a 
specific form, corresponding to torus fibrations over a base manifold. 
These are the geometries relevant for double field theory.


\subsection{Single doubled dimension}
\label{sec_rv_gen}

Let us start by applying the formalism explained in the previous section
to the setting of Ro\v cek and Verlinde \cite{Rocek:1991ps}.


\subsubsection*{Background}

More specifically, the RV background is given by first considering 
the following two-dimensional non-linear sigma-model
\begin{equation}\label{wsaction4}
\mathcal S^{\prime} ={1 \over 2\pi \alpha^{\prime}} \int dz \op d \bar{z} \left(
\op \Bigl[ \op g_{ab}(X) + b_{ab}(X) \op \Bigr] \partial X^a \bar{\partial} X^b 
+\frac{\alpha'}{4} R\, \phi(X) \op
\right)
\, ,
\end{equation}
where $a,b=3,\ldots, D+1$. Note that here we employed a complex basis on the world-sheet. 
In this action, the target space is $(D-1)$-dimensional with coordinates $X^a$.
Next, we enlarge the target space by introducing two additional coordinates $X_L$ and $X_R$, 
and define a new $(D+1)$-dimensional \emph{parent} sigma-model as
\eq{
&\mathcal S_{\rm LR}  =  \mathcal S^{\prime} + {1 \over 2 \pi \alpha^{\prime}} \int dz \op d \bar{z} \op \Big[ \:
\partial X_L \bar{\partial} X_L + \partial X_R \bar{\partial} X_R +
2 \op B(X) \op \partial X_R \bar{\partial} X_L  \\
&\hspace{155pt}+ 2\op  G_a^L (X) \op \partial X^a\bar{\partial} X_L +
2\op G_a^R (X) \op \bar{\partial} X^a\partial X_R \:\Big] \,,
\label{wsaction9}
}
whose couplings $B(X)$, $G_a^L(X)$ and $G_a^R(X)$ depend only on the $(D-1)$ coordinates $X^a$.
Note that this action is consistent with the action (\ref{action_01}). By comparison with the general 
form of a sigma-model, we see that
\eqref{wsaction9} determines a $(D+1)$-dimensional metric and $B$-field as follows
\eq{
  \label{rv_original}
  G_{IJ} = \left( \begin{array}{ccc}
  1 & B & G_b^L \\[4pt] 
  B & 1 & G_b^R \\[4pt]
  G_a^L & G_{a}^R & g_{ab}
  \end{array}\right) ,
  \hspace{35pt}
  B_{IJ} = \left( \begin{array}{ccc}
  0 & -B & -G_{b}^L \\[4pt] 
  +B & 0 & +G_{b}^R \\[4pt] 
  +G_{a}^L & -G_{a}^R & b_{ab}
  \end{array}\right) .
}
We then perform the field redefinitions 
\begin{equation}\label{basischange}
{ X}^1 = { X}_L + {X}_R \,, 
\hspace{70pt}
{ X}^2 = { X}_L - { X}_R \,,
\end{equation}
and  $G_{a}^\pm=G_{a}^L\pm G_{a}^R$,
after which the parent background \eqref{rv_original} can be expressed as
\eq{
  \label{rv_001}
  G_{IJ} = \left( \begin{array}{ccc}
  \frac{1+B}{2} & 0 & \frac{1}{2}\op G_{b}^+ \\[4pt] 
  0 & \frac{1-B}{2} & \frac{1}{2}\op G_{ b}^- \\[4pt]
  \frac{1}{2} \op G_{a}^+ & \frac{1}{2} \op G_{a}^- & g_{ab}
  \end{array}\right) ,
  \hspace{19pt}
  B_{IJ} = \left( \begin{array}{ccc}
  0 & +\frac{1}{2}\op B & -\frac{1}{2}\op G_{b}^- \\[4pt] 
  -\frac{1}{2}\op B & 0 & -\frac{1}{2}\op G_{b}^+ \\[4pt] 
  +\frac{1}{2}\op G_{a}^- & +\frac{1}{2} \op G_{a}^+ & b_{ab}
  \end{array}\right) .
 }
We also mention that compared to the previous section, we have slightly changed our notation for the indices. 
In the following we use
\eq{
  I,J = 1, \ldots, D+1 \,, 
  \hspace{30pt}
  i,j = 2, \ldots, D+1 \,, 
  \hspace{30pt}
  a,b = 3, \ldots, D+1 \,.
}


\subsubsection*{Isometries}

In \cite{Rocek:1991ps}, and also here,  the quantities $B$, $G_{a}^{\pm}$, $g_{ab}$ and 
$b_{ab}$ only depend on the coordinates $X^{a}$. Therefore, 
the metric $G_{IJ}$ in \eqref{rv_001} has at least two abelian isometries
generated by the Killing vectors
\eq{
  \label{iso_009}
  \arraycolsep2pt
  k^I_{(1)} = \left( \begin{array}{c} 1 \\ 0 \\  0 \end{array}\right)
  \hspace{50pt}
  {\rm and}
  \hspace{50pt}
  k^I_{(2)} = \left( \begin{array}{c} 0 \\ 1 \\  0 \end{array}\right).
}
Note that $G_{IJ}$ and $B_{IJ}$ do not constitute the most general $(D+1)$-dimensional sigma-model background, but are of rather restricted form.
For a general background with two abelian isometries of the form (\ref{iso_009}), the sigma-model action would be invariant under $O(2)\times O(2)$ transformations.
However, in order to preserve the restricted choice of the background as given in equations (\ref{rv_original}) and (\ref{rv_001}), 
the action is only invariant under a ${\mathbb Z}_2$ transformation. 
This can be  formulated as an automorphism
\begin{equation}
{{ X}}_L  \,\, \longrightarrow\,\,     \tilde{{ X}}_L = +{ X}_L  \, ,\hspace{50pt}  
{{ X}}_R  \,\, \longrightarrow\,\,     \tilde{{ X}}_R  = - { X}_R  \,,
\label{simosi}
\end{equation}
or in terms of the coordinates \eqref{basischange} as
\begin{equation}
{ X}^1={ X}_L+{ X}_R\quad  \longleftrightarrow\quad  { X}^2={ X}_L-{ X}_R\,,
\end{equation}
together will the following transformation on the background fields:
\begin{equation}\label{actiononB}
B\,\, \longleftrightarrow\,\,  -B\, ,
\hspace{40pt}
G_{a}^\pm\,\, \longleftrightarrow\,\,G_{a}^\mp\, .
\end{equation}
As we will see below, for the reduced background this transformation
is nothing else then standard
T-duality.


\subsubsection*{Reduced background}

Let us now follow the procedure explained in the previous section. 
In particular, we gauge a symmetry of the sigma-model by the following linear combination of the 
isometries \eqref{iso_009}
\eq{
  \label{rv_002}
  \arraycolsep2pt
  k^I = \left( \begin{array}{c} \mathsf a \\ \mathsf b \\  0 \end{array}\right),
  \hspace{70pt} \mathsf a^2 + \mathsf b^2 \neq0 \,.
}
Next, we determine the one-form $\mathsf v$ defined via \eqref{iso_02}. 
Solving the constraints \eqref{more_cons}, we obtain
\eq{
  \label{rv_005}
  \mathsf v = \frac{\mathsf a}{2}\, \Bigl[ (1-B)\op dX^2 + G^-_{\alpha} \op dX^{\alpha} \Bigr]
  + \frac{\mathsf b}{2}\, \Bigl[ (1+B) \op dX^1 + G^+_{\alpha} \op dX^{\alpha} \Bigr] \,.
}
However, imposing furthermore the condition \eqref{req_03} leads to $ \mathsf a \, \mathsf b =0$
which leaves only the two cases
\eq{
  \label{choices}
  (\mathsf a \neq0, \mathsf b = 0) \,, \hspace{70pt}
  (\mathsf a = 0, \mathsf b \neq 0)\,.
}
Note that here and in the following we assume that the field strength $H=dB$ is
non-zero along the directions $I=1,2$, that is $H_{12a}\neq0$. For vanishing $H$,
the restriction \eqref{choices} does not apply.

With the above data, we can now determine the $(D+1)$-dimensional 
matrix $\check G$ defined in \eqref{g_and_h_01}.
Since this matrix  has a null-eigenvector, 
we perform a change of coordinates such that the  reduced metric $\check{\mathcal G}$
only depends on the following $D$-dimensional basis of the co-tangent space
\eq{
  dY^2 = \mathsf b\op dX^1 - \mathsf a\op dX^2 \,, \hspace{60pt}
  dY^{a} = dX^{a} \,.
}
The dependence of the fields can be relabelled as $X^{a} = Y^{a}$.
With $i,j=2,\ldots, D+1$ we determine  the components of the reduced metric as 
\eq{
  \label{red_b_001}
  \check{\mathcal G}_{ij} = 
  \left( \begin{array}{@{\hspace{4pt}}c@{\hspace{16pt}}c@{\hspace{4pt}}}
  \displaystyle \frac{1\mp B}{1\pm B}
  &  \displaystyle  -\frac{G^{\mp}_{b}}{1\pm B} \\[12pt]
     \displaystyle -\frac{G^{\mp}_{a}}{1\pm B}
  &   \displaystyle g_{ab} \mp \frac{G_{a}^+G_{b}^+ - G_{a}^-G_{b}^-}{2\, (1\pm B)}
  \end{array}
  \right) ,
}
where the upper sign corresponds to the first choice in \eqref{choices} and the lower
sign to the second. 
Performing the same procedure for the $H$-field $\check H$ given in \eqref{g_and_h_01}, 
we can infer the corresponding reduced $B$-field as
\eq{
  \label{red_b_002}
  \check{\mathcal B}_{ij} = 
  \left( \begin{array}{@{\hspace{4pt}}c@{\hspace{16pt}}c@{\hspace{4pt}}}
  0 & \displaystyle \pm \frac{1}{1\pm B} \, G_{b}^{\pm}  \\[12pt]
  \displaystyle \mp \frac{1}{1\pm B} \, G_{a}^{\pm} &
  \displaystyle b_{ab} \mp \frac{G_{a}^{-}G_{b}^+ - G_{a}^+G_{b}^-}{2\, (1\pm B)}
  \end{array}
  \right) 
  .
}
Since these two theories descend from a single {\em parent} theory, in the following we 
call them {\em child} theories. 
As it is well-known, the two backgrounds specified by the upper and lower sign are
T-dual to each other. The transformations between them is given by the Buscher rules
 \cite{Buscher:1987sk,Buscher:1987qj,Buscher:1985kb}.


\subsubsection*{Discussion}

In this subsection, we have related the results of Ro\v cek and Verlinde in \cite{Rocek:1991ps} to our 
discussion in section~\ref{sec_dim_red}.
We made the following observations:
\begin{itemize}

\item First, for the choice of metric and $B$-field shown in \eqref{rv_original},
we have reproduced the two reduced backgrounds of  \cite{Rocek:1991ps} specified in 
\eqref{red_b_001} and \eqref{red_b_002}.

\item Second, since the parent background has the two abelian isometries \eqref{iso_009},
one might have expected that any linear combination \eqref{rv_002} can be gauged. 
This would give rise to continuous family of reduced backgrounds. 
However, for non-vanishing $H$-flux along the $I=1$ and $I=2$ direction
the constraints \eqref{cond_sum} allow only for two solutions, which 
explains the results of \cite{Rocek:1991ps} in a broader context.

\item Third, for the RV background we saw that at the level of the parent theory, 
T-duality corresponds to two different choices of Killing vectors. In view of our discussion 
in section~\ref{sec_dim_red}, this means choosing either $k$ or $v$ for the 
gauging procedure.

\end{itemize}


\subsection{Conformal invariance of parent and child theories}

In this section we  compare the $\beta$-functions of the $(D+1)$-dimensional parent theory with the
$\beta$-functions of the $D$-dimensional child theory.
In general the $\beta$-functions are given by the following set of equations
\eq{
\label{betafunctions}
\arraycolsep2pt
\begin{array}{@{}lcl@{}}
\beta^G_{IJ} &=& \displaystyle \alpha' \op R_{IJ} + 2\op \alpha' \op \nabla_I\nabla_J \Phi -\frac{\alpha'}{4}\op 
  H_{IKL} H_{J}{}^{KL} +\mathcal O(\alpha'^2)\,,  \\[8pt]
\beta^B_{IJ} &=& \displaystyle  - \frac{\alpha'}{2}\op \nabla^K H_{KIJ}+ \alpha'\op\nabla^K\Phi\, H_{KIJ} 
+\mathcal O(\alpha'^2) \,, \\[8pt]
\beta^{\Phi} &=& \displaystyle  \frac{D+1-D_{\rm crit}}{4} -\frac{\alpha'}{2}\op \nabla^2 \Phi+ \alpha' \op\nabla^K\Phi \nabla_K \Phi
- \frac{\alpha'}{24} \op H_{IJK}H^{IJK} +\mathcal O(\alpha'^2)\,.
\end{array}
}


\subsubsection*{$\beta$-functionals for a three-dimensional parent theory}

In order to keep our discussion tractable we focus on  $D=2$, that is we consider a three-dimensional
parent theory. In this case the 
$H$-field has only one non-trivial component, namely $H_{123}=\frac{1}{2}\op \partial_3B(X^3) = \frac{1}{2} B'$.
The vanishing of the $\beta$-functional for $B$ then reduces to 
\eq{
  0 =  H_{123}^{\prime}  - \Gamma_{3 \mu}^{\mu} \, H_{123} 
  - 2\op \Phi' H_{123} \,,
  \hspace{55pt}
  \partial_1\Phi=\partial_2\Phi = 0\,.
}
Note that here and in the remainder of this section, we denote the derivative with respect to $X^3$ by a prime.
Since  $\Gamma_{3 \mu}^{\mu} = \frac{1}{2} \log(\det G)^{\prime}$, it follows by integration that the dilaton is given by
\eq{
  \label{beta_01}
\Phi (X^3) = \phi_0 + {1 \over 4}  \log \left[{(H_{123})^2 \over \det G}\right] ,
}
with $\phi_0$ a constant. For this solution of the dilaton, 
all $\beta^B$- and $\beta^G$-functionals in three dimensions are vanishing, except
\eq{
  \label{beta_02}
  \beta^G_{33} = \alpha' \left[ \,  2\op \Phi'' -  \frac{(\det G)'}{\det G}\op \Phi' \,\right].
}
Let us now combine equation \eqref{beta_01} and the requirement of vanishing 
$\beta$-functional \eqref{beta_02}. We can infer two classes of solutions:
\begin{itemize}

\item[a)] The first class of models is characterized by a constant dilaton $\Phi$,
which trivially satisfy  $\beta^G_{33} =0$.
The flat torus in three dimensions as well as the $SU(2)$ and $SL(2, \mathbb{R})$ models are 
examples thereof. Note also that a constant dilaton implies
\eq{
  \frac{H_{123}}{\sqrt{\det G}} = \frac{\frac{1}{2} \op B'}{\sqrt{\det G}} = \gamma_1 = {\rm const.}
}

\item[b)] The second class of models is given by a non-constant dilaton. The vanishing 
of $\beta^G_{33}$ then implies that
\eq{
  \frac{\Phi'}{\sqrt{\det G}} = \gamma_{2} = {\rm const.} \neq 0\,.
}

\end{itemize}
We also mention that for the class of solutions a), the $\beta^{\Phi}$-functional (up to linear order in
$\alpha'$) is constant. 
Similarly, one can show in a somewhat more involved way that $\beta^{\Phi}$ is constant also for 
case b).


\subsubsection*{$\beta$-functionals for the two-dimensional child theories}

Let us now turn to the two-dimensional child theories. 
The metric and $B$-field are given by the expressions shown in \eqref{red_b_001} and \eqref{red_b_002}, and the 
dilaton of the child theories $\check \Phi$ is obtained by the usual Buscher shift, namely
\eq{
\check \Phi = \Phi - \frac{1}{2} \op\log(1 \pm B) \,.
}
For the $\beta$-functionals of the child theories we observe that in two-dimensions the $H$-field is vanishing, 
and hence the $\beta^B$-functional vanishes identically,
\eq{
  \check\beta^B_{ij} = 0 \,.
}
For $\check\beta^G$ we can perform an explicit computation
using for instance a computer algebra program. With the dilaton of the parent theory given 
in \eqref{beta_01}, we find that the $\beta^G$-functional of the child theories vanish, that is
\eq{
  \check\beta^G_{ij} = 0 \,.
}
The $\beta^{\Phi}$-functional for the child theories are computed slightly differently for the 
cases a) and b) mentioned above. However, for both we find that $\check\beta^{\Phi}$ is constant.
Moreover, the contribution $\Delta^{(1)}$ at linear order in $\alpha'$ is the same for the parent and child theories, i.e.
\eq{
  \label{red_dil}
  \arraycolsep2pt
  \begin{array}[c]{@{}lccc@{}}
  \beta^{\Phi} &=& \displaystyle \frac{D+1-D_{\rm crit}}{4} &+\,\alpha'  \Delta^{(1)}  + \mathcal O(\alpha'^2) \,, 
  \\[12pt]
  \displaystyle \check\beta^{\Phi} &=& \displaystyle \frac{D-D_{\rm crit}}{4} &+\,\alpha'  \Delta^{(1)}  
  + \mathcal O(\alpha'^2) \,. \\
  \end{array}
}
Note that the tree-level contribution to this $\beta$-functional is different for the parent and child theories. 
In view of their applications to T-duality, we require the child theories to be string theories, that is
$D=D_{\rm crit}$. In turn, this implies that the $\beta^{\Phi}$-functional of the parent theory is non-vanishing 
and hence the parent theory is not a string theory.


\subsection{Multiple doubled dimensions}
\label{sec_mdd}

Let us now generalize our discussion from section~\ref{sec_rv_gen} to multiple 
doubled dimensions. Here, we focus on the abelian case; the non-abelian situation 
will be discussed elsewhere. 


\subsubsection*{Doubled background}

We begin by specifying the geometry of the doubled background. We start again from 
the sigma-model \eqref{wsaction4}, and enlarge the target space by $2N$ coordinates
$\{X^{\alpha}_L,X^{\alpha}_R\}$.
The resulting action reads
\eq{
\mathcal S_{\rm LR}  =  \mathcal S^{\prime} + {1 \over 2 \pi \alpha^{\prime}} \int dz \op d \bar{z} \op \Big[ \,
\delta_{\alpha\beta}\op \partial X^{\alpha}_L \bar{\partial} X^{\beta}_L + 
\delta_{\alpha\beta}\op \partial X^{\alpha}_R \bar{\partial} X^{\beta}_R +
2 \op B_{\alpha\beta} \op \partial X^{\alpha}_R \bar{\partial} X^{\beta}_L&  \\
+ 2\op  G_{a\beta}^L  \op \partial X^a\bar{\partial} X^{\beta}_L +
2\op G_{a\beta}^R \op \bar{\partial} X^a\partial X^{\beta}_R& \,\Big] , \hspace{-4pt}
\label{wsaction11}
}
where $\alpha,\beta=1,\ldots, N$ and $a,b=2N+1, \ldots, 2N+D$,
and where we assume all couplings $B_{\alpha\beta}$, $G^{L/R}_{a\beta}$ to only depend
on the coordinates $X^a$.
Next, we perform a change of coordinates 
\eq{
  X^{\alpha} = X_{L}^{\alpha} + X_R^{\alpha} \,, 
  \hspace{70pt}
  X^{N+\alpha} = X_{L}^{\alpha} - X_R^{\alpha} \,.
}
and define
\eq{
  \arraycolsep2pt
  \begin{array}{lcl@{\hspace{60pt}}lcl}
  B^+_{\alpha\beta} &=& \displaystyle \tfrac{1}{2} \bigl(B_{\alpha\beta}+B^T_{\alpha\beta} \bigr) \,, 
  &
  G^+_{a\beta} = G^L_{a\beta} + G^R_{a\beta} \,, 
  \\[6pt]
  B^-_{\alpha\beta} &=& \displaystyle \tfrac{1}{2} \bigl(B_{\alpha\beta}-B^T_{\alpha\beta} \bigr) \,,
  &
  G^-_{a\beta} = G^L_{a\beta} - G^R_{a\beta} \,.  
  \end{array}
}
Suppressing the indices, the resulting metric and $B$-field (written in a basis $X^I=\{X^{\alpha}, X^{N+\alpha}, X^a\}$) take the following form
\eq{
  \label{rv_001_a}
  \arraycolsep3pt
  G_{IJ} = \left( \begin{array}{ccc}
  \frac{\mathds 1+B^+}{2} & +\frac{1}{2}\op B^- & \frac{1}{2}\op G^{+\op T} \\[6pt] 
  -\frac{1}{2}\op B^- & \frac{\mathds 1-B^+}{2} & \frac{1}{2}\op G^{-\op T} \\[6pt]
  \frac{1}{2} \op G^+ & \frac{1}{2} \op G^- & g
  \end{array}\right) \!,
  \hspace{13pt}
  B_{IJ} = \left( \begin{array}{ccc}
  +\frac{1}{2} \op B^- & +\frac{1}{2}\op B^+ & -\frac{1}{2}\op G^{-\op T} \\[6pt] 
  -\frac{1}{2}\op B^+ & -\frac{1}{2}\op B^- & -\frac{1}{2}\op G^{+\op T} \\[6pt] 
  +\frac{1}{2}\op G^- & +\frac{1}{2} \op G^+ & b
  \end{array}\right) \!.
 }
Note that since all components only depend on $X^a$, this background has at least $2N$ isometries.
Without loss of generality, we can express the corresponding Killing vectors as
\eq{
  \label{kv_009}
  k_{\alpha}^I =
  \scalebox{0.8}{$\displaystyle
  \renewcommand{\arraystretch}{1.1}
  \left( \begin{array}{c}\vdots \\ 1 \\ \vdots \\ \hdashline[2pt/1.5pt] \\ 0 \\ \\ \hline
  0 \end{array}\right) $} 
  = \delta^I_{\alpha} \,,
  \hspace{80pt}
  v_{\alpha}^I =
  \scalebox{0.8}{$\displaystyle
  \renewcommand{\arraystretch}{1.1}
  \left( \begin{array}{c} \\ 0 \\ \\ \hdashline[2pt/1.5pt] \vdots \\ 1 \\ \vdots \\ \hline
  0 \end{array}\right) $} 
  = \delta^I_{N+\alpha} \,.
}
Furthermore, the background specified in \eqref{rv_001_a} is invariant under the following
combined transformation
\eq{
  X^{\alpha} \; \longleftrightarrow \; X^{N+\alpha} \,, \hspace{40pt}
  G^+_{a\beta} \; \longleftrightarrow \; G^-_{a\beta} \,, \hspace{40pt}
  B_{\alpha\beta} \; \longleftrightarrow \; - B_{\alpha\beta}\,,
}
which implies $k_{\alpha} \leftrightarrow v_{\alpha}$ for the Killing vectors.


\subsubsection*{Reduced background}

Let us now follow the procedure introduced in section~\ref{sec_dim_red} to obtain 
the reduced background. In particular,
we want to gauge the $N$ isometries specified by $k_{\alpha}$ 
and integrate-out the corresponding gauge fields. 

The vectors \eqref{kv_009} satisfy the conditions shown in \eqref{cond_sum} and \eqref{cond_sum_2}.
We can therefore apply the reduction procedure, 
for which the matrices $\check G$ and $\check B$ defined in \eqref{red_geo} take the general form
\eq{
  \label{red_back_001}
  \arraycolsep5pt
  \renewcommand{\arraystretch}{1.5}
  \check G_{IJ} = \left( \begin{array}{c:c@{\hspace{3pt}}c}
  0 & 0 & 0 \\ \hdashline
  0 & \check G^{(22)}  & \check G^{(23)} \\
  0 & \check G^{(32)} & \check G^{(33)}
  \end{array}\right),
  \hspace{50pt}
  \arraycolsep5pt
  \renewcommand{\arraystretch}{1.5}
  \check B_{IJ} = \left( \begin{array}{c:c@{\hspace{3pt}}c}
  0 & 0 & 0 \\ \hdashline
  0 & \check B^{(22)}  & \check B^{(23)} \\
  0 & \check B^{(32)} & \check B^{(33)}
  \end{array}\right),
 }
where we applied a gauge transformation to remove a constant term in $\check B$. 
The explicit expressions for the components can easily be determined from \eqref{red_geo}. 
However, to illustrate the underlying structure we consider the case $B^-=0$ for which 
we find
\eq{
  \label{red_093}
  \arraycolsep2pt
  \begin{array}{lcl}
  \check G^{(22)} &=& \displaystyle \mathds 1 \, \bigl( 1+ B^+ \bigr)^{-1}\bigl( 1- B^+ \bigr) \,, \\[6pt]
  \check G^{(23)} &=& \displaystyle +\mathds 1 \, \bigl( 1+ B^+ \bigr)^{-1}\, G^{-\op T} \,, \\[6pt]
  \check G^{(33)} &=& \displaystyle g - \tfrac{1}{2}\op G^+ \bigl( 1+ B^+ \bigr)^{-1} G^{+\op T}
    + \tfrac{1}{2}\op G^- \bigl( 1+ B^+ \bigr)^{-1} G^{-\op T} \,,
  \\[16pt]
  \check B^{(22)} &=& \displaystyle 0 \,, \\[6pt]
  \check B^{(23)} &=&- \displaystyle \mathds 1\, \bigl( 1+ B^+ \bigr)^{-1}\, G^{+\op T} \,, \\[6pt]
   \check B^{(33)} &=& \displaystyle b - \tfrac{1}{2}\op G^- \bigl( 1+ B^+ \bigr)^{-1} G^{+\op T}
    + \tfrac{1}{2}\op G^+ \bigl( 1+ B^+ \bigr)^{-1} G^{-\op T} \,.
  \end{array}
}
These are the generalizations of \eqref{red_b_001} and \eqref{red_b_002}.
Note that if we choose to perform the reduction by gauging the vectors $v_{\alpha}$ 
instead of $k_{\alpha}$, we obtain \eqref{red_093} with the replacements $B^+\to - B^+$ and
$G^{\pm} \to G^{\mp}$.
We have therefore shown, that our analysis of a single doubled dimensions
generalizes to multiple doubled dimensions. 
In particular, T-duality is again given by interchanging the vector fields 
$k_{\alpha}$ and $v_{\alpha}$.


\section{Examples}
\label{sec_ex}

Let us now apply the procedure introduced in the previous sections  to
Wess-Zumino-Witten models. We first consider the $SU(2)$ example,
and then comment on the generalization to arbitrary Lie groups.
Note that WZW models in the context of DFT have also been studied in 
\cite{Blumenhagen:2014gva,Bosque:2015jda}, 
albeit in a rather different approach.


\subsection{$SU(2)$ WZW model}
\label{sec_su2}

The $SU(2)$ WZW model corresponds to a conformal field theory on a three-sphere with 
non-vanishing $H$-flux. We therefore consider a three-dimensional parent 
theory and want to construct two-dimensional child theories.


\subsubsection*{Parent theory}

We start by introducing the setting. We choose the following parametrization for the metric of a 
round three-sphere with radius $R$
\eq{
  \label{su2_met}
  G_{IJ} = R^2 \left(\begin{array}{ccc}
  \sin^2 \eta & 0 & 0 \\
  0 & \cos^2\eta & 0 \\
  0 & 0 & 1 
  \end{array} \right) .
}
The  coordinates are given by $x^I = (\zeta_1, \zeta_2, \eta)$ with $\zeta_{1,2}\in [0,2\pi)$ 
and $\eta\in [0,\pi/2]$.
In order for the three-dimensional model to be conformal,  the $H$-flux and dilaton
have to take the form
\eq{  
  \label{su2_h}
  H = 2\op R^2 \sin \eta\cos\eta\: d\zeta_1 \wedge d\zeta_2 \wedge d\eta \,,
  \hspace{60pt}
  \Phi= {\rm const}.
}
And, indeed, for this background the one-loop $\beta$-functionals for the metric and $B$-field shown in 
\eqref{betafunctions} are vanishing.
Note also that this background can be brought into the form \eqref{rv_001}. 
In particular, if we identify
\eq{
  \label{connect_2_rv}
  \arraycolsep2pt
  \begin{array}{lcl@{\hspace{50pt}}lcl@{\hspace{50pt}}lcl}
  X^1 &=& \displaystyle R\, \zeta_1 \,, & B &=& -\cos(2\eta) \,, &G^+_3 &=& 0 \,,\\[10pt]
  X^2 &=& \displaystyle R\, \zeta_2 \,, & g_{33} &=& R^2 \,, &G^-_3 &=& 0 \,, \\[10pt]
  X^3 &=& \displaystyle \eta \,, &
  \end{array}
}
the metric shown in \eqref{rv_001} becomes \eqref{su2_met}, and the field strength computed 
from $B_{IJ}$ in  \eqref{rv_001} becomes \eqref{su2_h}.


\subsubsection*{Child theories}

Next, this background has two abelian Killing vectors given by  
$k^I=(1,0,0)^T$ and $k^I=(0,1,0)^T$. In order to solve the constraints 
\eqref{cond_sum}, we saw in section~\ref{sec_rv_gen} that not every linear combination can be
used for the reduction procedure but that only two 
choices are possible. In particular, because the metric and $B$-field of the three-sphere background can be 
brought into the RV form, \eqref{choices} implies that 
the two allowed gaugings are
\eq{
  \label{sphere_kv}
  \arraycolsep2pt
  k_{(1)}^I = \left( \begin{array}{c} 1 \\ 0 \\  0 \end{array}\right),
  \hspace{70pt} 
  k_{(2)}^I = \left( \begin{array}{c} 0 \\ 1 \\  0 \end{array}\right).
}
We remark that the overall normalization of the Killing vectors drops out of all formulas. 
With the help of the identifications \eqref{connect_2_rv}, we can now determine the child theories from the general 
expressions \eqref{red_b_001} and \eqref{red_b_002}. 
\begin{enumerate}

\item For the first Killing vector in \eqref{sphere_kv}, we obtain the following reduced background
\eq{
  \arraycolsep4pt
  \check{\mathcal G}_{ij} = R^2 \left(\begin{array}{cc}
  \cot^2\eta & 0  \\[6pt]
  0 & 1 
  \end{array} \right) ,
  \hspace{60pt}
  \check{\mathcal B}_{ij} = 0\,.
}
This geometry is known as the trumpet solution, 
which is a conformal model provided the 
dilaton is adjusted through the Buscher shift as
\eq{
  \check \Phi =  \Phi - \log(\sin\eta) \,.
}

\item For the second Killing vector shown in equation \eqref{sphere_kv}, we obtain the reduced background
as the cigar solution
\eq{
  \arraycolsep4pt
  \check{\mathcal G}_{ij} = R^2 \left(\begin{array}{cc}
  \tan^2\eta & 0  \\[6pt]
  0 & 1 
  \end{array} \right) ,
  \hspace{60pt}
  \check{\mathcal B}_{ij} = 0\,.
}
This is again a conformal model, if we adjust the reduced dilaton as
\eq{
  \check \Phi =  \Phi - \log(\cos\eta) \,.
}

\end{enumerate}
For the $SU(2)$ example we have therefore verified our above results, namely that 
for the RV-type background two isometries can be gauged.
The resulting child theories are conformal, and it is well-known that they are T-dual to each other.
In particular, T-duality corresponds to interchanging which Killing vector has 
been used to perform the reduction.


\subsection{General WZW model}

In this subsection we now comment on the 
generalization of the reduction procedure to arbitrary Lie groups. 
We first review some properties of Wess-Zumino-Witten models, 
and then discuss the gauging constraints \eqref{cond_sum} and 
\eqref{cond_sum_2}.


\subsubsection*{Action}

Let us start with  a WZW model for a Lie group $G$.
The corresponding action (without the dilaton term) is given by 
\eq{
  \label{wzw_action}
  \mathcal S =\hphantom{+\,}&\frac{1}{2\pi\alpha'}\int_{\partial\Sigma} {\rm Tr}\op\Bigl[\,
  \tfrac{k}{4} \op g^{-1} dg \wedge \star g^{-1}dg \,\Bigr] \\
  +\,& \frac{i}{2\pi\alpha'}\int_{\Sigma} \hspace{4.9pt}{\rm Tr}\op\Bigl[\,
  \tfrac{k}{6} \op g^{-1} dg \wedge g^{-1} dg \wedge g^{-1} dg \,\Bigr] \,,
}
where $g\in G$ and where $k$ denotes the level. The generators 
of the corresponding Lie algebra
$\{t_{\mathsf a}\}\in \mathfrak g$ 
with $\mathsf a=1,\ldots,D$
are normalized such that 
\eq{
  \bigl[\op t_{\mathsf a}, t_{\mathsf b} \op\bigr] = i\op f_{\mathsf{ab}}{}^{\mathsf c} \op t_{\mathsf c} \,,
  \hspace{70pt}
  {\rm Tr}\op \bigl( t_{\mathsf a} \op t_{\mathsf b} \bigr) = 2\op \delta_{\mathsf{ab}} \,,
}
and the left- and right-invariant forms are given by
\eq{
  \label{lrinvforms}
  \omega_L = g^{-1} dg = \omega_L{}^{\mathsf a} \op t_{\mathsf a}\,, \hspace{70pt}
  \omega_R = dg \op g^{-1}=\omega_R{}^{\mathsf a}\op t_{\mathsf a} \,.
}
Using these conventions, we can determine the target-space metric $G$ and the field strength $H$
by comparing with \eqref{action_01}
as follows
\eq{
  \label{wzw_back}
  \arraycolsep2pt
  \begin{array}{lcc@{\hspace{1pt}}cclcc@{\hspace{1pt}}ccl}
  G &=& -&k & \delta_{\mathsf {ab}} &\omega_L{}^{\mathsf a} \wedge \star \omega_L{}^{\mathsf b} 
  &=& -&k & \delta_{\mathsf {ab}} &\omega_R{}^{\mathsf a} \wedge \star \omega_R{}^{\mathsf b}\,, \\[8pt]
  H&=& -&\displaystyle\frac{i\op k}{3!} &f_{\mathsf{abc}} &\omega_L{}^{\mathsf a} \wedge \omega_L{}^{\mathsf b} 
     \wedge \omega_L{}^{\mathsf c}
  &=& -&\displaystyle \frac{i\op k}{3!} & f_{\mathsf{abc}} & \omega_R{}^{\mathsf a} \wedge \omega_R{}^{\mathsf b} 
     \wedge \omega_R{}^{\mathsf c}
  \,.
  \end{array}
}
Here, $f_{\mathsf{abc}} = f_{\mathsf{ab}}{}^{\mathsf d}\op\delta_{\mathsf {dc}}$ is completely 
anti-symmetric in its indices, and the minus sign in the metric ensures that $G_{\mathsf{ab}}$ is positive definite. 
Indeed, for $g$ unitary 
it follows that $(g^{-1} dg)^{\dagger} = - d^{-1} dg$ and thus the first line in \eqref{wzw_action}
is negative definite.


\subsubsection*{Geometry}

Let us now briefly recall some of the properties of the target-space geometry. 
First, the left- and right-invariant forms \eqref{lrinvforms} satisfy the Maurer-Cartan 
equation, that is
\eq{
  \label{wzw_003}
  0 = d\op \omega_L{}^{\mathsf a} + \frac{i}{2}\op f_{\mathsf{bc}}{}^{\mathsf a} \omega_L{}^{\mathsf b}
    \wedge \omega_L{}^{\mathsf c} \,, 
  \hspace{50pt}
  0 = d\op \omega_R{}^{\mathsf a} - \frac{i}{2}\op f_{\mathsf{bc}}{}^{\mathsf a} \omega_R{}^{\mathsf b}
    \wedge \omega_R{}^{\mathsf c} \,.
}
Next, we choose a coordinate basis of one-forms $\{dX^i\}$ with $i=1,\ldots,D$, 
and expand \eqref{lrinvforms} as
\eq{
  \omega_L{}^a = \omega_L{}^{\mathsf a}{}_i \, dX^i \,,
  \hspace{70pt}
  \omega_R{}^a = \omega_R{}^{\mathsf a}{}_i \, dX^i \,.
}
The dual vector fields $\xi_{L\op\mathsf a}$ and $\xi_{R\op\mathsf a}$ are defined via 
$\iota_{\xi_{\mathsf a}}\omega^{\mathsf b} = \delta_{\mathsf a}{}^{\mathsf b}$ for the left- and right-invariant 
sector, leading to the expressions
\eq{
  \xi_{L\op \mathsf a}{} = (\omega_L^{-1})^i{}_{\mathsf a} \, \partial_i \,, 
  \hspace{70pt}
  \xi_{R\op \mathsf a}{} = (\omega_R^{-1})^i{}_{\mathsf a} \, \partial_i \,.
}
They are the left- and right-invariant vector fields. For later reference, we furthermore define
\eq{
  \label{wzw_004}
  \iota_{\xi_{L\mathsf a}} \omega_R{}^{\mathsf b} 
  = \omega_R{}^{\mathsf b}{}_i \op(\omega^{-1}_L)^i{}_{\mathsf a} =   R^{\mathsf b}{}_{\mathsf a} \,,
  \hspace{40pt}
  \iota_{\xi_{R\mathsf a}} \omega_L{}^{\mathsf b} 
  = \omega_L{}^{\mathsf b}{}_i \op(\omega^{-1}_R)^i{}_{\mathsf a}    = (R^{-1})^{\mathsf b}{}_{\mathsf a} \,.
}
The vector fields satisfy the following algebra
\eq{
  \arraycolsep2pt
  \begin{array}{ccl}
  \displaystyle \bigl[ \op \xi_{L\op \mathsf a} , \xi_{L\op \mathsf b} \op \bigr] &=& 
  \displaystyle + i \op f_{\mathsf {ab}}{}^{\mathsf c}  \xi_{L\op \mathsf c}  \,, 
  \\[10pt]
  \displaystyle \bigl[ \op \xi_{R\op \mathsf a} , \xi_{R\op \mathsf b} \op \bigr] &=& 
  \displaystyle - i \op f_{\mathsf {ab}}{}^{\mathsf c} \xi_{R\op \mathsf c}  \,, 
  \end{array}
  \hspace{70pt}
  \bigl[ \op \xi_{L\op \mathsf a} , \xi_{R\op \mathsf b} \op \bigr] = 0 \,,
}
and they are Killing vectors for the metric $G$ shown in \eqref{wzw_back}, that is
\eq{
  \mathcal L_{\xi_{L\op \mathsf a}} G = 0\,,
  \hspace{70pt}
  \mathcal L_{\xi_{R\op \mathsf a}} G = 0\,.
}
This means, that the isometry group for the Lie group $G$ is $G_L\times G_R$.


\subsubsection*{Gauging conditions I}

We now want to construct vector fields $k_{\alpha}$ and $v_{\alpha}$ which satisfy 
the gauging conditions summarized in equations \eqref{cond_sum} and
\eqref{cond_sum_2}.
We make the following choice
\eq{
  \label{wzw_001}
  k_{\mathsf a} = -i\op \bigl( \op \xi_{L\op\mathsf a} -  \xi_{R\op\mathsf a}  \bigr) \,,
  \hspace{60pt}
  v_{\mathsf a} = +i\op \bigl( \op \xi_{L\op\mathsf a} +  \xi_{R\op\mathsf a}  \bigr) \,,
}
for which the dual one-forms $\mathsf k_{\mathsf a} = k_{\mathsf a}^i \op G_{ij} \op dX^j$
and $\mathsf v_{\mathsf a} = v_{\mathsf a}^i \op G_{ij} \op dX^j$ are given by
\eq{
  \label{wzw_002}
  \mathsf k_{\mathsf a} = +i\op k \, \delta_{\mathsf{ab}}\bigl( \op \omega_L{}^{\mathsf b} -  
    \omega_R{}^{\mathsf b} \op \bigr) \,,
  \hspace{60pt}
  \mathsf v_{\mathsf a} = -i\op k \,  \delta_{\mathsf{ab}}\bigl( \op \omega_L{}^{\mathsf b} +  
    \omega_R{}^{\mathsf b} \op \bigr) \,,
}
where $k$ denotes again the level.
Let us now discuss the implications for the gauging conditions:
\begin{itemize}

\item Using the properties listed above, we can show that \eqref{wzw_001} and \eqref{wzw_002}
satisfy the conditions shown in \eqref{cond_sum_2}.

\item Furthermore, since the left- and right-invariant vectors $\xi_{L\op\mathsf a}$
and $\xi_{R\op\mathsf a}$ are Killing vectors, also $k_{\mathsf a}$
and $v_{\mathsf a}$ are Killing. 
Moreover, also the second row in  \eqref{cond_sum}  is satisfied, which can be seen by we computing 
\eq{
  d \op \mathsf k_{\mathsf a} - \iota_{v_{\mathsf a}} H 
    =0 \,,
  \hspace{60pt}
  d \op \mathsf v_{\mathsf a} - \iota_{k_{\mathsf a}} H 
    = 0 \,.
}

\item Finally, we have to demand that the matrix $\mathcal G$ defined in \eqref{back_67} is invertible. 
Let us therefore compute
\eq{
  \mathcal G_{\mathsf{ab}} = k_{\mathsf a}^i \op G_{ij} \op k^j_{\mathsf b} = k \, \bigl[ \mathds 1 - R^T 
    \bigr]_{\mathsf a}^{\hspace{4pt}\mathsf c} \, \delta_{\mathsf{cd}} \,\bigl[ \mathds 1 - R 
    \bigr]^{\mathsf d}_{\hspace{4pt}\mathsf b} \,, 
}
where the matrix $R$ had been defined in \eqref{wzw_004}.
However,  $\mathcal G_{\mathsf{ab}}$ with $\mathsf a,\mathsf b= 1,\ldots, D$ 
always has an eigenvector with vanishing 
eigenvalue.\footnote{
Note that $R^{\mathsf a}{}_{\mathsf b} = \frac{1}{2}\op \delta^{\mathsf{ac}} \op {\rm Tr}(t_{\mathsf c}\op  g \op t_{\mathsf b}\op g^{-1})$, where $g\in G$. Writing then $g= \exp(\op i\op \phi^{\mathsf a}t_{\mathsf a})$ 
and using the Baker-Campell-Hausdorff formula, we see that $R^{\mathsf a}{}_{\mathsf b} \phi^{\mathsf b}
= \phi^{\mathsf a}$. Hence, $(\mathds 1 - R)^{\mathsf a}{}_{\mathsf b}$ has $\phi^{\mathsf a}$ as an
eigenvector with vanishing eigenvalue. 
}
Hence, the condition \eqref{invert_067} for invertibility of $\mathcal G$ is not satisfied and it is therefore not possible to gauge all 
isometries with Killing vectors $k_{\mathsf a}$
for $\mathsf a = 1,\ldots, D$.

\end{itemize}


\subsubsection*{Gauging conditions II}

However, one possibility to obtain a matrix $\mathcal G$ which satisfies \eqref{invert_067} 
is to not gauge all $k_{\mathsf a}$ for 
$\mathsf a = 1, \ldots, D$, but to choose a subset. For instance, let us consider
\eq{
  \label{wzw_006}
  \arraycolsep2pt
  \begin{array}{lcl@{\hspace{80pt}}lcl}
  k_{\alpha} &=& -i\op \bigl( \op \xi_{L\op\alpha} -  \xi_{R\op\alpha}  \bigr) \,,
  & \xi_{L\op\alpha}&\in&{\rm Cartan}(\mathfrak g_L) \,,
  \\[6pt]
  v_{\alpha} &=& +i\op \bigl( \op \xi_{L\op\alpha} +  \xi_{R\op\alpha}  \bigr) \,,
  & \xi_{R\op\alpha}&\in&{\rm Cartan}(\mathfrak g_R) \,,
  \end{array}
}
where the left- and right-invariant vector fields are elements of the Cartan subalgebra 
$\mathfrak g_L$ and $\mathfrak g_R$ and where $\alpha=1,\ldots, N$
with $N<D$.
Since $[\mathfrak g_L,\mathfrak g_R]=0$, this implies that all $k_{\alpha}$ and $v_{\alpha}$ 
commute among each other
\eq{
  \bigl[ k_{\alpha} , k_{\beta} \bigr]_{\rm L}  =0 \,,
  \hspace{50pt}
  \bigl[ k_{\alpha} , v_{\beta} \bigr]_{\rm L}  =0  \,,
  \hspace{50pt}
  \bigl[ v_{\alpha} ,  v_{\beta} \bigr]_{\rm L} =0 \,.   
}
Furthermore, the matrix $\mathcal G_{\alpha\beta}$ defined in \eqref{back_67} 
is in general invertible, except at singular points. This is precisely the setting we have encountered 
in section~\ref{sec_su2} for the $SU(2)$ case.


\subsubsection*{Discussion}

In this subsection we have seen, that the example of the $SU(2)$ WZW model can 
be generalized to arbitrary Lie groups. For the choice of 
Killing vectors \eqref{wzw_002} the gauging conditions 
\eqref{cond_sum} and  \eqref{cond_sum_2} can be satisfied, however, 
the matrix $\mathcal G_{\mathsf{ab}}$ is not invertible when gauging all $k_{\mathsf a}$-isometries.

On the other hand, when choosing only a subset of isometries, the matrix $\mathcal G_{\alpha\beta}$ is invertible except at special points. 
A convenient subset of isometries corresponds to the Cartan algebra, shown in \eqref{wzw_006}.
It would be interesting to further investigate this point and to construct explicit models, 
however, this is beyond the scope of this paper.


\section{Summary and conclusions}
\label{sec_sc}

In this paper, we have discussed how through a gauging procedure 
the number of target-space dimensions of a 
non-linear sigma-model can be reduced.
In particular, if the parent theory  exhibits target-space isometries satisfying
the conditions \eqref{cond_sum} and \eqref{cond_sum_2}, it is possible to gauge
a corresponding world-sheet symmetry and integrate-out the gauge field. 
As we have shown, the resulting theory corresponds to a 
background which is reduced by one dimension. 
We have then generalized this procedure to gauging $N$
symmetries, resulting in reductions by $N$ dimensions.

In section~\ref{sec_rv} we have considered a particular class of geometries
relevant for double field theory. In the context of T-duality, these have 
appeared in \cite{Rocek:1991ps} and have also been used in \cite{Bakas:2015gia}.
They are characterized by the presence of an even number of abelian isometries, and can 
be viewed as torus fibrations over a base manifold. 
We have shown that these backgrounds satisfy the gauging constraints summarized in 
equations \eqref{cond_sum} and \eqref{cond_sum_2}, and we 
have performed the corresponding reduction. 
Different choices for which isometries are used for the reduction lead to different 
child theories, which are T-dual to each other.

For the three-dimensional situation we have furthermore illustrated, that if the parent background is 
conformal the child theories are conformal as well. 
As shown in equation \eqref{red_dil}, the only difference occurs for the dilaton, 
where the corresponding $\beta$-functional of the parent theory is non-vanishing at tree-level 
due to not being a critical string theory. The reduced child theories on the other
hand are indeed string theories with vanishing $\beta$-functionals.

In section~\ref{sec_ex} we have discussed the $SU(2)$ WZW model as an example. 
This model is conformal and corresponds to the three-sphere with $H$-flux, and the isometry group
contains two abelian isometries. Applying the gauging procedure for 
one of these two isometries, leads to a conformal model in two dimensions. 
As it turns out, it is not possible to choose any linear combination of 
these isometries, but only two choices satisfy the constraints 
\eqref{cond_sum} and \eqref{cond_sum_2}. These lead to the 
cigar and trumpet solution in two dimensions, which are again conformal models and 
T-dual to each other. 
We have furthermore commented on the generalization to  WZW models on arbitrary 
Lie groups.

Our studies in this paper provide a starting point for the development of 
a world-sheet description of double field theory. 
The novel feature of our formalism is that the reduction from the  parent 
theory to the child theories is realized through a gauging procedure,
instead of imposing constraints explicitly. This gauging procedure should correspond to imposing the strong constraint in double field theory, which also
eliminates half of the coordinates.
The next task in this program is to relate the $\beta$-functionals of the world-sheet theory
to the equations of motion of double field theory, which we plan to address in 
future work. 
Other open questions are to find further explicit examples where the proposed gauging
reduction is realized. In particular, it would be interesting to study 
examples with non-abelian isometry groups. 
We hope to come back to these points in the future.


\vskip1.25em
\subsubsection*{Acknowledgements}

We thank Christoph Mayrhofer and Cornelius Schmidt-Colinet for helpful discussions. 
EP would like to thank the Physics Department at Seoul National University 
and the Erwin Schr\"odinger International Institute in Vienna
for hospitality. This work was partially supported by the ERC Advanced Grant "Strings and Gravity"
(Grant.No. 32004) and by the DFG cluster of excellence "Origin and Structure of the Universe".


\clearpage
\begin{appendix}

\section{Requirement of vanishing field strength}
\label{app_da}

In this appendix we give some details on the computation leading to 
the constraints \eqref{more_cons} and \eqref{more_cons_a}. However, since
the abelian case is included in the non-abelian situation, we focus on the latter.

When gauging a sigma-model with non-abelian isometries, we impose the constraint
\eqref{da_is_zero} in order to not increase the number of degrees of freedom of the theory. 
As the matrix $\mathcal M_{\alpha\beta}$ defined in \eqref{back_67} is assumed to be 
invertible, we can write \eqref{da_is_zero} also as
\eq{
  \label{app_002}
  0 = \mathcal M_{\alpha\beta} \op F^{\beta} \,.
}
To compute $F^{\alpha}$, we need to know how the exterior derivative acts on $\star dX^i$ on 
the world-sheet. This is determined by the equation of motion of the gauged action \eqref{action_02}
for $X^i$ which read
\eq{
  0 = &-G_{ij} \Bigl[ \,\Gamma^j_{pq} \,dX^p\wedge\star dX^q + d\star dX^j \, \Bigr] \\
  & + \bigl( \iota_i d\mathsf k_{\alpha} \bigr) \wedge\star A^{\alpha} 
  - \mathsf k_{\alpha\op i} \, d\star A^{\alpha}
  + \frac{1}{2}\op \partial_i \mathcal G_{\alpha\beta} \, A^{\alpha} \wedge \star A^{\beta} \\
  &+ i \, \iota_i H + i \bigl( \iota_i d\mathsf v_{\alpha} \bigr) \wedge A^{\alpha}
  - \frac{i}{2}\, \bigl( \iota_i \iota_{k_{\alpha}} \iota_{k_{\beta}} H \bigr) A^{\alpha} \wedge A^{\beta} \\
  & + \frac{\alpha'}{2} \op R\, \partial_i \Phi \star 1 \,,
}
where the Christoffel symbols $\Gamma_{ij}^k$ are computed using the target-space metric $G_{ij}$.
Furthermore, it is useful to note that from \eqref{req_002} we can derive the following relation
\eq{
  0 = 3 \, \iota_{k_{[\ul \alpha}} f_{\ul \beta \ul \gamma]}{}^{\epsilon}\op  \mathsf v_{\epsilon}
  - \iota_{k_{\alpha}}\iota_{k_{\beta}}\iota_{k_{\gamma}} H \,,
}
which implies 
\eq{
  \mathcal L_{k_{\alpha}} \mathcal G_{\beta\gamma}= f_{\alpha\beta}{}^{\epsilon} \op\mathcal G_{\epsilon \gamma}
      + f_{\alpha\gamma}{}^{\epsilon} \op\mathcal G_{\epsilon \beta} \,,
  \hspace{40pt}
  \mathcal L_{k_{\alpha}} \mathcal D_{\beta\gamma}= f_{\alpha\beta}{}^{\epsilon} \op\mathcal D_{\epsilon \gamma}- 
      f_{\alpha\gamma}{}^{\epsilon} \op\mathcal D_{\epsilon \beta}\,.
}      
For \eqref{app_002} we  then compute
\eq{
  \label{app_003}
   0 =&
   \quad \mathcal M_{\alpha\beta} \op F^{\beta} \\
   =& 
  \arraycolsep1.5pt
   \begin{array}[t]{@{}crclrccrccrl@{}}
   &\multicolumn{8}{l}{\displaystyle \frac{\alpha'}{2} \op R \op \bigl[ \, v_{\alpha}^m \partial_m \Phi \bigr]\star 1} \\[12pt]
   +& \Bigl[  & A_{\alpha \op ij} & DX^i \wedge & \star DX^{j} &
   -2 \op B_{\alpha\op ij} & DX^i \wedge & \star (k\op \omega)^j &
   - A_{\alpha\op ij} & (k\op \omega)^i \wedge & \star (k\op \omega)^j &\Bigr] \\[12pt]
   - & i \op \Bigl[  & B_{\alpha \op ij} & DX^i \wedge  & DX^{j} &
   -2 \op A_{\alpha\op ij} & DX^i \wedge & (k\op \omega)^j &
   - B_{\alpha\op ij} & (k\op \omega)^i \wedge & (k\op \omega)^j & \Bigr]   \,,
   \end{array}
}
where we have defined
\eq{
  \arraycolsep2pt
  \begin{array}{lcl@{\hspace{50pt}}lcl}
  DX^i &=& dX^i - k^i_{\mu} \, l^{\mu} \,, &
  l^{\mu} &=& \mathcal M^{\mu\nu} \big( \mathsf k + \mathcal D \op \mathcal G^{-1}\, \mathsf v\bigr)_{\nu}
  \,, 
  \\[8pt]
  (k\omega)^i &=& k^i_{\mu} \, \omega^{\mu} \,, &
  \omega^{\mu} &=& \mathcal M^{\mu\nu} \big( \mathsf v + \mathcal D \op \mathcal G^{-1}\, \mathsf k\bigr)_{\nu}
  \,,
  \end{array}
}  
as well as
\eq{
   A_{\alpha\op ij} = \nabla_{( \op \ov i} \op \mathsf v_{\alpha\op \ov j \op)}  \,, 
   \hspace{50pt}
   B_{\alpha\op ij} = \nabla_{[ \op \ul i} \op \mathsf k_{\alpha\op \ul j \op]} - \frac{1}{2} \op v_{\alpha}^m H_{mij}  \,.
}
Equation \eqref{app_003} can be split into a real and imaginary part at zeroth order in $\alpha'$, and
in a part linear in $\alpha'$. Assuming $\alpha'$ to be an expansion parameter, we infer 
from the latter part that $v_{\alpha}^m\partial_m \Phi = 0$. 
Concerning the real and imaginary part, we have not been able to show that $F^{\alpha}=0$ implies
that $A_{\alpha\op ij}=0$ and $B_{\alpha\op ij}=0$. 
This is related to the fact that when written as matrix equations, $A_{\alpha}$ and $B_{\alpha}$ have $k^i_{\mu}$ as eigenvectors with vanishing eigenvalue.
Furthermore, \eqref{app_003} is invariant under the following transformation
\eq{
  \arraycolsep2.2pt
  \begin{array}{lclcccccc}
  A_{\alpha\op ij} &\to& A_{\alpha\op ij} &+& a_{\alpha}{}^{\beta}{}_{(\ov i} & \mathsf k_{\beta\op \ov j)}
  &+& b_{\alpha}{}^{\beta}{}_{(\ov i} & \mathsf v_{\beta\op \ov j)} \,, \\[6pt]
  B_{\alpha\op ij} &\to& B_{\alpha\op ij} &+& b_{\alpha}{}^{\beta}{}_{[\ul i} & \mathsf k_{\beta\op \ul j]}
  &+& a_{\alpha}{}^{\beta}{}_{[\ul i} & \mathsf v_{\beta\op \ul j]} \,, \\
  \end{array}
}
where $a_{\alpha}{}^{\beta}{}_i$ and $b_{\alpha}{}^{\beta}{}_i$ are arbitrary functions. 
However, it is of course true that setting $A_{\alpha\op ij}$ and $B_{\alpha\op ij}$ to zero 
(and imposing $v_{\alpha}^m\partial_m \Phi = 0$) solves $F^{\alpha}=0$.
This is what we assume for this paper, namely that for all $\alpha$
\eq{
   A_{\alpha\op ij} =0 \,,  \hspace{50pt}
   B_{\alpha\op ij} = 0 \,,  \hspace{50pt}
   v_{\alpha}^m\partial_m \Phi = 0 \,,
}
which corresponds to \eqref{more_cons} and \eqref{more_cons_a} in the main text.


\end{appendix}


\clearpage
\bibliography{references}  

\providecommand{\href}[2]{#2}\begingroup\raggedright\begin{thebibliography}{10}

\bibitem{Witten:1995ex}
E.~Witten, ``{String theory dynamics in various dimensions},'' {\em Nucl.
  Phys.} {\bf B443} (1995) 85--126,
\href{http://www.arXiv.org/abs/hep-th/9503124}{{\tt hep-th/9503124}}.

\bibitem{Giveon:1994fu}
A.~Giveon, M.~Porrati, and E.~Rabinovici, ``{Target space duality in string
  theory},'' {\em Phys. Rept.} {\bf 244} (1994) 77--202,
\href{http://www.arXiv.org/abs/hep-th/9401139}{{\tt hep-th/9401139}}.

\bibitem{Buscher:1985kb}
T.~H. Buscher, ``{Quantum Corrections and Extended Supersymmetry in New
  $\sigma$ Models},'' {\em Phys. Lett.} {\bf B159} (1985)
127.

\bibitem{Buscher:1987sk}
T.~H. Buscher, ``{A Symmetry of the String Background Field Equations},'' {\em
  Phys. Lett.} {\bf B194} (1987)
59.

\bibitem{Buscher:1987qj}
T.~H. Buscher, ``{Path Integral Derivation of Quantum Duality in Nonlinear
  Sigma Models},'' {\em Phys. Lett.} {\bf B201} (1988)
466.

\bibitem{Rocek:1991ps}
M.~Rocek and E.~P. Verlinde, ``{Duality, quotients, and currents},'' {\em Nucl.
  Phys.} {\bf B373} (1992) 630--646,
\href{http://www.arXiv.org/abs/hep-th/9110053}{{\tt hep-th/9110053}}.

\bibitem{Giveon:1988tt}
A.~Giveon, E.~Rabinovici, and G.~Veneziano, ``{Duality in String Background
  Space},'' {\em Nucl. Phys.} {\bf B322} (1989)
167.

\bibitem{Giveon:1991jj}
A.~Giveon and M.~Rocek, ``{Generalized duality in curved string backgrounds},''
  {\em Nucl. Phys.} {\bf B380} (1992) 128--146,
\href{http://www.arXiv.org/abs/hep-th/9112070}{{\tt hep-th/9112070}}.

\bibitem{Giveon:1993ph}
A.~Giveon and E.~Kiritsis, ``{Axial vector duality as a gauge symmetry and
  topology change in string theory},'' {\em Nucl. Phys.} {\bf B411} (1994)
  487--508,
\href{http://www.arXiv.org/abs/hep-th/9303016}{{\tt hep-th/9303016}}.

\bibitem{Alvarez:1993qi}
E.~Alvarez, L.~Alvarez-Gaume, J.~L.~F. Barbon, and Y.~Lozano, ``{Some global
  aspects of duality in string theory},'' {\em Nucl. Phys.} {\bf B415} (1994)
  71--100,
\href{http://www.arXiv.org/abs/hep-th/9309039}{{\tt hep-th/9309039}}.

\bibitem{Alvarez:1994wj}
E.~Alvarez, L.~Alvarez-Gaume, and Y.~Lozano, ``{A Canonical approach to duality
  transformations},'' {\em Phys. Lett.} {\bf B336} (1994) 183--189,
\href{http://www.arXiv.org/abs/hep-th/9406206}{{\tt hep-th/9406206}}.

\bibitem{Bakas:1995hc}
I.~Bakas and K.~Sfetsos, ``{T duality and world sheet supersymmetry},'' {\em
  Phys. Lett.} {\bf B349} (1995) 448--457,
\href{http://www.arXiv.org/abs/hep-th/9502065}{{\tt hep-th/9502065}}.

\bibitem{Plauschinn:2013wta}
E.~Plauschinn, ``{T-duality revisited},'' {\em JHEP} {\bf 01} (2014) 131,
\href{http://www.arXiv.org/abs/1310.4194}{{\tt 1310.4194}}.

\bibitem{Plauschinn:2014nha}
E.~Plauschinn, ``{On T-duality transformations for the three-sphere},'' {\em
  Nucl. Phys.} {\bf B893} (2015) 257--286,
\href{http://www.arXiv.org/abs/1408.1715}{{\tt 1408.1715}}.

\bibitem{Chatzistavrakidis:2015lga}
A.~Chatzistavrakidis, A.~Deser, and L.~Jonke, ``{T-duality without isometry via
  extended gauge symmetries of 2D sigma models},''
\href{http://www.arXiv.org/abs/1509.01829}{{\tt 1509.01829}}.

\bibitem{Dasgupta:1999ss}
K.~Dasgupta, G.~Rajesh, and S.~Sethi, ``{M theory, orientifolds and G -
  flux},'' {\em JHEP} {\bf 08} (1999) 023,
\href{http://www.arXiv.org/abs/hep-th/9908088}{{\tt hep-th/9908088}}.

\bibitem{Kachru:2002sk}
S.~Kachru, M.~B. Schulz, P.~K. Tripathy, and S.~P. Trivedi, ``{New
  supersymmetric string compactifications},'' {\em JHEP} {\bf 03} (2003) 061,
\href{http://www.arXiv.org/abs/hep-th/0211182}{{\tt hep-th/0211182}}.

\bibitem{Hellerman:2002ax}
S.~Hellerman, J.~McGreevy, and B.~Williams, ``{Geometric constructions of
  nongeometric string theories},'' {\em JHEP} {\bf 01} (2004) 024,
\href{http://www.arXiv.org/abs/hep-th/0208174}{{\tt hep-th/0208174}}.

\bibitem{Dabholkar:2002sy}
A.~Dabholkar and C.~Hull, ``{Duality twists, orbifolds, and fluxes},'' {\em
  JHEP} {\bf 09} (2003) 054,
\href{http://www.arXiv.org/abs/hep-th/0210209}{{\tt hep-th/0210209}}.

\bibitem{Hull:2004in}
C.~M. Hull, ``{A Geometry for non-geometric string backgrounds},'' {\em JHEP}
  {\bf 10} (2005) 065,
\href{http://www.arXiv.org/abs/hep-th/0406102}{{\tt hep-th/0406102}}.

\bibitem{Shelton:2005cf}
J.~Shelton, W.~Taylor, and B.~Wecht, ``{Nongeometric flux compactifications},''
  {\em JHEP} {\bf 10} (2005) 085,
\href{http://www.arXiv.org/abs/hep-th/0508133}{{\tt hep-th/0508133}}.

\bibitem{Mathai:2004qq}
V.~Mathai and J.~M. Rosenberg, ``{T duality for torus bundles with H fluxes via
  noncommutative topology},'' {\em Commun. Math. Phys.} {\bf 253} (2004)
  705--721,
\href{http://www.arXiv.org/abs/hep-th/0401168}{{\tt hep-th/0401168}}.

\bibitem{Mathai:2004qc}
V.~Mathai and J.~M. Rosenberg, ``{On Mysteriously missing T-duals, H-flux and
  the T-duality group},'' in {\em {Differential geometry and physics.
  Proceedings, 23rd International Conference, Tianjin, China, August 20-26,
  2005}}, pp.~350--358.
\newblock 2004.
\newblock
\href{http://www.arXiv.org/abs/hep-th/0409073}{{\tt hep-th/0409073}}.
\newblock

\bibitem{Bouwknegt:2004ap}
P.~Bouwknegt, K.~Hannabuss, and V.~Mathai, ``{Nonassociative tori and
  applications to T-duality},'' {\em Commun. Math. Phys.} {\bf 264} (2006)
  41--69,
\href{http://www.arXiv.org/abs/hep-th/0412092}{{\tt hep-th/0412092}}.

\bibitem{Bouwknegt:2004tr}
P.~Bouwknegt, K.~Hannabuss, and V.~Mathai, ``{T-duality for principal torus
  bundles and dimensionally reduced Gysin sequences},'' {\em Adv. Theor. Math.
  Phys.} {\bf 9} (2005), no.~5, 749--773,
\href{http://www.arXiv.org/abs/hep-th/0412268}{{\tt hep-th/0412268}}.

\bibitem{Ellwood:2006my}
I.~Ellwood and A.~Hashimoto, ``{Effective descriptions of branes on
  non-geometric tori},'' {\em JHEP} {\bf 12} (2006) 025,
\href{http://www.arXiv.org/abs/hep-th/0607135}{{\tt hep-th/0607135}}.

\bibitem{Grange:2006es}
P.~Grange and S.~Schafer-Nameki, ``{T-duality with H-flux: Non-commutativity,
  T-folds and G x G structure},'' {\em Nucl. Phys.} {\bf B770} (2007) 123--144,
\href{http://www.arXiv.org/abs/hep-th/0609084}{{\tt hep-th/0609084}}.

\bibitem{Blumenhagen:2010hj}
R.~Blumenhagen and E.~Plauschinn, ``{Nonassociative Gravity in String
  Theory?},'' {\em J. Phys.} {\bf A44} (2011) 015401,
\href{http://www.arXiv.org/abs/1010.1263}{{\tt 1010.1263}}.

\bibitem{Lust:2010iy}
D.~L{\"u}st, ``{T-duality and closed string non-commutative (doubled)
  geometry},'' {\em JHEP} {\bf 12} (2010) 084,
\href{http://www.arXiv.org/abs/1010.1361}{{\tt 1010.1361}}.

\bibitem{Blumenhagen:2011ph}
R.~Blumenhagen, A.~Deser, D.~L{\"u}st, E.~Plauschinn, and F.~Rennecke,
  ``{Non-geometric Fluxes, Asymmetric Strings and Nonassociative Geometry},''
  {\em J. Phys.} {\bf A44} (2011) 385401,
\href{http://www.arXiv.org/abs/1106.0316}{{\tt 1106.0316}}.

\bibitem{Blumenhagen:2011yv}
R.~Blumenhagen, ``{Nonassociativity in String Theory},'' in {\em Strings, gauge
  fields, and the geometry behind: The legacy of Maximilian Kreuzer},
  A.~Rebhan, L.~Katzarkov, J.~Knapp, R.~Rashkov, and E.~Scheidegger, eds.,
  pp.~213--224.
\newblock World Scientific, 2011.
\newblock
\href{http://www.arXiv.org/abs/1112.4611}{{\tt 1112.4611}}.
\newblock

\bibitem{Lust:2012fp}
D.~L{\"u}st, ``{Twisted Poisson Structures and Non-commutative/non-associative
  Closed String Geometry},'' {\em PoS} {\bf CORFU2011} (2011) 086,
\href{http://www.arXiv.org/abs/1205.0100}{{\tt 1205.0100}}.

\bibitem{Plauschinn:2012kd}
E.~Plauschinn, ``{Non-geometric fluxes and non-associative geometry},'' {\em
  PoS} {\bf CORFU2011} (2011) 061,
\href{http://www.arXiv.org/abs/1203.6203}{{\tt 1203.6203}}.

\bibitem{Condeescu:2012sp}
C.~Condeescu, I.~Florakis, and D.~L{\"u}st, ``{Asymmetric Orbifolds,
  Non-Geometric Fluxes and Non-Commutativity in Closed String Theory},'' {\em
  JHEP} {\bf 04} (2012) 121,
\href{http://www.arXiv.org/abs/1202.6366}{{\tt 1202.6366}}.

\bibitem{Mylonas:2012pg}
D.~Mylonas, P.~Schupp, and R.~J. Szabo, ``{Membrane Sigma-Models and
  Quantization of Non-Geometric Flux Backgrounds},'' {\em JHEP} {\bf 09} (2012)
  012,
\href{http://www.arXiv.org/abs/1207.0926}{{\tt 1207.0926}}.

\bibitem{Chatzistavrakidis:2012qj}
A.~Chatzistavrakidis and L.~Jonke, ``{Matrix theory origins of non-geometric
  fluxes},'' {\em JHEP} {\bf 02} (2013) 040,
\href{http://www.arXiv.org/abs/1207.6412}{{\tt 1207.6412}}.

\bibitem{Andriot:2012vb}
D.~Andriot, M.~Larfors, D.~L{\"u}st, and P.~Patalong, ``{(Non-)commutative
  closed string on T-dual toroidal backgrounds},'' {\em JHEP} {\bf 06} (2013)
  021,
\href{http://www.arXiv.org/abs/1211.6437}{{\tt 1211.6437}}.

\bibitem{Bakas:2013jwa}
I.~Bakas and D.~L{\"u}st, ``{3-Cocycles, Non-Associative Star-Products and the
  Magnetic Paradigm of R-Flux String Vacua},'' {\em JHEP} {\bf 01} (2014) 171,
\href{http://www.arXiv.org/abs/1309.3172}{{\tt 1309.3172}}.

\bibitem{Deser:2013pra}
A.~Deser, ``{Lie algebroids, non-associative structures and non-geometric
  fluxes},'' {\em Fortsch. Phys.} {\bf 61} (2013) 1056--1153,
\href{http://www.arXiv.org/abs/1309.5792}{{\tt 1309.5792}}.

\bibitem{Mylonas:2013jha}
D.~Mylonas, P.~Schupp, and R.~J. Szabo, ``{Non-Geometric Fluxes, Quasi-Hopf
  Twist Deformations and Nonassociative Quantum Mechanics},'' {\em J. Math.
  Phys.} {\bf 55} (2014) 122301,
\href{http://www.arXiv.org/abs/1312.1621}{{\tt 1312.1621}}.

\bibitem{Blair:2014kla}
C.~D.~A. Blair, ``{Non-commutativity and non-associativity of the doubled
  string in non-geometric backgrounds},'' {\em JHEP} {\bf 06} (2015) 091,
\href{http://www.arXiv.org/abs/1405.2283}{{\tt 1405.2283}}.

\bibitem{Mylonas:2014aga}
D.~Mylonas, P.~Schupp, and R.~J. Szabo, ``{Nonassociative geometry and twist
  deformations in non-geometric string theory},'' {\em PoS} {\bf ICMP2013}
  (2013) 007,
\href{http://www.arXiv.org/abs/1402.7306}{{\tt 1402.7306}}.

\bibitem{Aschieri:2015roa}
P.~Aschieri and R.~J. Szabo, ``{Triproducts, nonassociative star products and
  geometry of R-flux string compactifications},'' {\em J. Phys. Conf. Ser.}
  {\bf 634} (2015), no.~1, 012004,
\href{http://www.arXiv.org/abs/1504.03915}{{\tt 1504.03915}}.

\bibitem{Kawai:1986ah}
H.~Kawai, D.~C. Lewellen, and S.~H.~H. Tye, ``{Construction of Fermionic String
  Models in Four-Dimensions},'' {\em Nucl. Phys.} {\bf B288} (1987)
1.

\bibitem{Lerche:1986cx}
W.~Lerche, D.~L{\"u}st, and A.~N. Schellekens, ``{Chiral Four-Dimensional
  Heterotic Strings from Selfdual Lattices},'' {\em Nucl. Phys.} {\bf B287}
  (1987)
477.

\bibitem{Antoniadis:1986rn}
I.~Antoniadis, C.~P. Bachas, and C.~Kounnas, ``{Four-Dimensional
  Superstrings},'' {\em Nucl. Phys.} {\bf B289} (1987)
87.

\bibitem{Narain:1986qm}
K.~S. Narain, M.~H. Sarmadi, and C.~Vafa, ``{Asymmetric Orbifolds},'' {\em
  Nucl. Phys.} {\bf B288} (1987)
551.

\bibitem{Flournoy:2005xe}
A.~Flournoy and B.~Williams, ``{Nongeometry, duality twists, and the
  worldsheet},'' {\em JHEP} {\bf 01} (2006) 166,
\href{http://www.arXiv.org/abs/hep-th/0511126}{{\tt hep-th/0511126}}.

\bibitem{Halmagyi:2008dr}
N.~Halmagyi, ``{Non-geometric String Backgrounds and Worldsheet Algebras},''
  {\em JHEP} {\bf 07} (2008) 137,
\href{http://www.arXiv.org/abs/0805.4571}{{\tt 0805.4571}}.

\bibitem{Halmagyi:2009te}
N.~Halmagyi, ``{Non-geometric Backgrounds and the First Order String Sigma
  Model},''
\href{http://www.arXiv.org/abs/0906.2891}{{\tt 0906.2891}}.

\bibitem{Rennecke:2014sca}
F.~Rennecke, ``{O(d,d)-Duality in String Theory},'' {\em JHEP} {\bf 10} (2014)
  69,
\href{http://www.arXiv.org/abs/1404.0912}{{\tt 1404.0912}}.

\bibitem{Bakas:2015gia}
I.~Bakas and D.~L{\"u}st, ``{T-duality, Quotients and Currents for
  Non-Geometric Closed Strings},'' {\em Fortsch. Phys.} {\bf 63} (2015)
  543--570,
\href{http://www.arXiv.org/abs/1505.04004}{{\tt 1505.04004}}.

\bibitem{Chatzistavrakidis:2015vka}
A.~Chatzistavrakidis, L.~Jonke, and O.~Lechtenfeld, ``{Sigma models for
  genuinely non-geometric backgrounds},'' {\em JHEP} {\bf 11} (2015) 182,
\href{http://www.arXiv.org/abs/1505.05457}{{\tt 1505.05457}}.

\bibitem{Andriot:2011uh}
D.~Andriot, M.~Larfors, D.~L{\"u}st, and P.~Patalong, ``{A ten-dimensional
  action for non-geometric fluxes},'' {\em JHEP} {\bf 09} (2011) 134,
\href{http://www.arXiv.org/abs/1106.4015}{{\tt 1106.4015}}.

\bibitem{Andriot:2012wx}
D.~Andriot, O.~Hohm, M.~Larfors, D.~L{\"u}st, and P.~Patalong, ``{A geometric
  action for non-geometric fluxes},'' {\em Phys. Rev. Lett.} {\bf 108} (2012)
  261602,
\href{http://www.arXiv.org/abs/1202.3060}{{\tt 1202.3060}}.

\bibitem{Andriot:2012an}
D.~Andriot, O.~Hohm, M.~Larfors, D.~L{\"u}st, and P.~Patalong, ``{Non-Geometric
  Fluxes in Supergravity and Double Field Theory},'' {\em Fortsch. Phys.} {\bf
  60} (2012) 1150--1186,
\href{http://www.arXiv.org/abs/1204.1979}{{\tt 1204.1979}}.

\bibitem{Blumenhagen:2012nk}
R.~Blumenhagen, A.~Deser, E.~Plauschinn, and F.~Rennecke, ``{A bi-invariant
  Einstein-Hilbert action for the non-geometric string},'' {\em Phys. Lett.}
  {\bf B720} (2013) 215--218,
\href{http://www.arXiv.org/abs/1210.1591}{{\tt 1210.1591}}.

\bibitem{Blumenhagen:2012nt}
R.~Blumenhagen, A.~Deser, E.~Plauschinn, and F.~Rennecke, ``{Non-geometric
  strings, symplectic gravity and differential geometry of Lie algebroids},''
  {\em JHEP} {\bf 02} (2013) 122,
\href{http://www.arXiv.org/abs/1211.0030}{{\tt 1211.0030}}.

\bibitem{Blumenhagen:2013aia}
R.~Blumenhagen, A.~Deser, E.~Plauschinn, F.~Rennecke, and C.~Schmid, ``{The
  Intriguing Structure of Non-geometric Frames in String Theory},'' {\em
  Fortsch. Phys.} {\bf 61} (2013) 893--925,
\href{http://www.arXiv.org/abs/1304.2784}{{\tt 1304.2784}}.

\bibitem{Andriot:2013xca}
D.~Andriot and A.~Betz, ``{$\beta$-supergravity: a ten-dimensional theory with
  non-geometric fluxes, and its geometric framework},'' {\em JHEP} {\bf 12}
  (2013) 083,
\href{http://www.arXiv.org/abs/1306.4381}{{\tt 1306.4381}}.

\bibitem{Andriot:2014uda}
D.~Andriot and A.~Betz, ``{NS-branes, source corrected Bianchi identities, and
  more on backgrounds with non-geometric fluxes},'' {\em JHEP} {\bf 07} (2014)
  059,
\href{http://www.arXiv.org/abs/1402.5972}{{\tt 1402.5972}}.

\bibitem{Tseytlin:1990nb}
A.~A. Tseytlin, ``{Duality Symmetric Formulation of String World Sheet
  Dynamics},'' {\em Phys. Lett.} {\bf B242} (1990)
163--174.

\bibitem{Tseytlin:1990va}
A.~A. Tseytlin, ``{Duality symmetric closed string theory and interacting
  chiral scalars},'' {\em Nucl. Phys.} {\bf B350} (1991)
395--440.

\bibitem{Duff:1989tf}
M.~J. Duff, ``{Duality Rotations in String Theory},'' {\em Nucl. Phys.} {\bf
  B335} (1990)
610.

\bibitem{Dall'Agata:2008qz}
G.~Dall'Agata and N.~Prezas, ``{Worldsheet theories for non-geometric string
  backgrounds},'' {\em JHEP} {\bf 08} (2008) 088,
\href{http://www.arXiv.org/abs/0806.2003}{{\tt 0806.2003}}.

\bibitem{Hohm:2013jaa}
O.~Hohm, W.~Siegel, and B.~Zwiebach, ``{Doubled $\alpha'$-geometry},'' {\em
  JHEP} {\bf 02} (2014) 065,
\href{http://www.arXiv.org/abs/1306.2970}{{\tt 1306.2970}}.

\bibitem{Nibbelink:2012jb}
S.~Groot~Nibbelink and P.~Patalong, ``{A Lorentz invariant doubled world-sheet
  theory},'' {\em Phys. Rev.} {\bf D87} (2013), no.~4, 041902,
\href{http://www.arXiv.org/abs/1207.6110}{{\tt 1207.6110}}.

\bibitem{Nibbelink:2013zda}
S.~Groot~Nibbelink, F.~Kurz, and P.~Patalong, ``{Renormalization of a Lorentz
  invariant doubled worldsheet theory},'' {\em JHEP} {\bf 10} (2014) 114,
\href{http://www.arXiv.org/abs/1308.4418}{{\tt 1308.4418}}.

\bibitem{Dabholkar:2005ve}
A.~Dabholkar and C.~Hull, ``{Generalised T-duality and non-geometric
  backgrounds},'' {\em JHEP} {\bf 05} (2006) 009,
\href{http://www.arXiv.org/abs/hep-th/0512005}{{\tt hep-th/0512005}}.

\bibitem{Hull:2006va}
C.~M. Hull, ``{Doubled Geometry and T-Folds},'' {\em JHEP} {\bf 07} (2007) 080,
\href{http://www.arXiv.org/abs/hep-th/0605149}{{\tt hep-th/0605149}}.

\bibitem{Siegel:1993xq}
W.~Siegel, ``{Two vierbein formalism for string inspired axionic gravity},''
  {\em Phys. Rev.} {\bf D47} (1993) 5453--5459,
\href{http://www.arXiv.org/abs/hep-th/9302036}{{\tt hep-th/9302036}}.

\bibitem{Siegel:1993th}
W.~Siegel, ``{Superspace duality in low-energy superstrings},'' {\em Phys.
  Rev.} {\bf D48} (1993) 2826--2837,
\href{http://www.arXiv.org/abs/hep-th/9305073}{{\tt hep-th/9305073}}.

\bibitem{Hull:2009mi}
C.~Hull and B.~Zwiebach, ``{Double Field Theory},'' {\em JHEP} {\bf 09} (2009)
  099,
\href{http://www.arXiv.org/abs/0904.4664}{{\tt 0904.4664}}.

\bibitem{Hohm:2010jy}
O.~Hohm, C.~Hull, and B.~Zwiebach, ``{Background independent action for double
  field theory},'' {\em JHEP} {\bf 07} (2010) 016,
\href{http://www.arXiv.org/abs/1003.5027}{{\tt 1003.5027}}.

\bibitem{Hohm:2010pp}
O.~Hohm, C.~Hull, and B.~Zwiebach, ``{Generalized metric formulation of double
  field theory},'' {\em JHEP} {\bf 08} (2010) 008,
\href{http://www.arXiv.org/abs/1006.4823}{{\tt 1006.4823}}.

\bibitem{Aldazabal:2013sca}
G.~Aldazabal, D.~Marques, and C.~Nunez, ``{Double Field Theory: A Pedagogical
  Review},'' {\em Class. Quant. Grav.} {\bf 30} (2013) 163001,
\href{http://www.arXiv.org/abs/1305.1907}{{\tt 1305.1907}}.

\bibitem{Hohm:2013bwa}
O.~Hohm, D.~L{\"u}st, and B.~Zwiebach, ``{The Spacetime of Double Field Theory:
  Review, Remarks, and Outlook},'' {\em Fortsch. Phys.} {\bf 61} (2013)
  926--966,
\href{http://www.arXiv.org/abs/1309.2977}{{\tt 1309.2977}}.

\bibitem{Hori:1999me}
K.~Hori, ``{D-branes, T duality, and index theory},'' {\em Adv. Theor. Math.
  Phys.} {\bf 3} (1999) 281--342,
\href{http://www.arXiv.org/abs/hep-th/9902102}{{\tt hep-th/9902102}}.

\bibitem{Bouwknegt:2003vb}
P.~Bouwknegt, J.~Evslin, and V.~Mathai, ``{T duality: Topology change from H
  flux},'' {\em Commun. Math. Phys.} {\bf 249} (2004) 383--415,
\href{http://www.arXiv.org/abs/hep-th/0306062}{{\tt hep-th/0306062}}.

\bibitem{Hitchin:2004ut}
N.~Hitchin, ``{Generalized Calabi-Yau manifolds},'' {\em Quart. J. Math.} {\bf
  54} (2003) 281--308,
\href{http://www.arXiv.org/abs/math/0209099}{{\tt math/0209099}}.

\bibitem{Gualtieri:2003dx}
M.~Gualtieri, {\em {Generalized complex geometry}}.
\newblock PhD thesis, Oxford U., 2003.
\newblock
\href{http://www.arXiv.org/abs/math/0401221}{{\tt math/0401221}}.
\newblock

\bibitem{Grana:2008yw}
M.~Grana, R.~Minasian, M.~Petrini, and D.~Waldram, ``{T-duality, Generalized
  Geometry and Non-Geometric Backgrounds},'' {\em JHEP} {\bf 04} (2009) 075,
\href{http://www.arXiv.org/abs/0807.4527}{{\tt 0807.4527}}.

\bibitem{Hull:1989jk}
C.~M. Hull and B.~J. Spence, ``{The Gauged Nonlinear $\sigma$ Model With
  {Wess-Zumino} Term},'' {\em Phys. Lett.} {\bf B232} (1989)
204.

\bibitem{Hull:1990ms}
C.~M. Hull and B.~J. Spence, ``{The Geometry of the gauged sigma model with
  Wess-Zumino term},'' {\em Nucl. Phys.} {\bf B353} (1991)
379--426.

\bibitem{Blumenhagen:2014gva}
R.~Blumenhagen, F.~Hassler, and D.~L{\"u}st, ``{Double Field Theory on Group
  Manifolds},'' {\em JHEP} {\bf 02} (2015) 001,
\href{http://www.arXiv.org/abs/1410.6374}{{\tt 1410.6374}}.

\bibitem{Bosque:2015jda}
P.~d. Bosque, F.~Hassler, and D.~L{\"u}st, ``{Flux Formulation of DFT on Group
  Manifolds and Generalized Scherk-Schwarz Compactifications},'' {\em JHEP}
  {\bf 02} (2016) 039,
\href{http://www.arXiv.org/abs/1509.04176}{{\tt 1509.04176}}.

\end{thebibliography}\endgroup
\bibliographystyle{utphys}


\end{document}